\documentclass[conf]{new-aiaa}
\usepackage[utf8]{inputenc}

\newcommand{\makeJG}[1]{#1}
\newcommand{\makeFG}[1]{#1}
\usepackage[dvipsnames]{xcolor}

\def \pp#1#2{\frac{\partial #1}{\partial #2}}

\newcommand{\bu}{\mathbf{u}}
\newcommand{\bn}{\mathbf{n}}

\newcommand{\code}{\textit{PyFlowCL}}

\def \Rey {\mathrm{Re}}
\def \Ma {\mathrm{Ma}}
\def \Pr {\mathrm{Pr}}

\newcommand{\bI}{{\mathbf{I}}}

\newcommand{\bQ}{{\mathbf{Q}}}
\newcommand{\bF}{{\mathbf{F}}}
\newcommand{\bR}{{\mathbf{R}}}

\usepackage{amsmath}

\usepackage{algorithm}
\usepackage{algpseudocode}
\newcommand{\etal}{\textit{et al.}}

\newcommand\figref{Figure~\ref}

\newcommand{\clcd}{C_l/C_d}
\newcommand{\clcdc}{(C_l/C_d)_c}
\usepackage{graphicx}
\usepackage{amsmath}
\usepackage{longtable,tabularx}
\usepackage{multirow}
\usepackage{xcolor}
\usepackage{booktabs}
\usepackage{caption}
\usepackage{subcaption}

\title{Active Control of Turbulent Airfoil Flows Using Adjoint-based Deep Learning}

\author{Xuemin Liu\footnote{Graduate Student, Department of Aerospace and Mechanical Engineering.}}
\affil{University of Notre Dame, Notre Dame, IN 46556, USA}

\author{Tom Hickling\footnote{Postdoctoral Research Associate, Mathematical Institute.}
}
\affil{University of Oxford, Oxford OX2 6GG, UK}

\author{Jonathan~F.~MacArt\footnote{Assistant Professor, Department of Aerospace and Mechanical Engineering, jmacart@nd.edu. Member AIAA. \\ Presented as Paper 2025-1300 at the 2025 AIAA SciTech Forum, Orlando, FL, January 6--10, 2025.}}
\affil{University of Notre Dame, Notre Dame, IN 46556, USA}

\begin{document}

\maketitle

\begin{abstract}

We train active neural-network flow controllers using a deep learning PDE augmentation method to optimize lift-to-drag ratios in turbulent airfoil flows at Reynolds number $5\times10^4$ and Mach number 0.4. Direct numerical simulation and large eddy simulation are employed to model compressible, unconfined flow over two- and three-dimensional semi-infinite NACA 0012 airfoils at angles of attack $\alpha = 5^\circ$, $10^\circ$, and $15^\circ$. 
Control actions, implemented through a blowing/suction jet at a fixed location and geometry on the upper surface, are adaptively determined by a neural network that maps local pressure measurements to optimal jet total pressure, enabling a sensor-informed control policy that responds spatially and temporally to unsteady flow conditions.
The sensitivities of the flow to the neural network parameters are computed using the adjoint Navier–Stokes equations, which we construct using automatic differentiation applied to the flow solver.
The trained flow controllers significantly improve the lift-to-drag ratios and reduce flow separation for both two- and three-dimensional airfoil flows, especially at $\alpha = 5^\circ$ and $10^\circ$. The 2D-trained models remain effective when applied out-of-sample to 3D flows, which demonstrates the robustness of the adjoint-trained control approach. The 3D-trained models capture the flow dynamics even more effectively, which leads to better energy efficiency and comparable performance for both adaptive (neural network) and offline (simplified, constant-pressure) controllers. These results underscore the effectiveness of this learning-based approach in improving aerodynamic performance.

\end{abstract}
\section{Introduction}

The need to increase the efficiency of air transport has become a paramount priority as demand continues to increase. Recent developments in active flow control have shown a potential to reduce fuel consumption and emissions by 20--25\,\%~\cite{Dussauge2023reinforcement, Skinner2018state}. 
Since drag is the primary source of energy loss in flight~\cite{Rabault2019artificial}, the ability to optimize airfoil flows---crucial for lift generation and aerodynamic efficiency---is essential. 
This need has led to an increased focus on improving the lift-to-drag ratio in airfoil flows through various flow control techniques~\cite{ozkan2022active,satar2024a}. 
Flow control techniques can be broadly categorized into passive and active control. 
Passive flow control (PFC) harnesses inherent flow dynamics to achieve control without requiring external energy input. In contrast, active flow control (AFC) involves active manipulation through energy addition and/or removal. Both PFC and AFC aim to significantly improve aerodynamic efficiency and thus achieve increased sustainability of air transportation.

Common PFC methods include geometric adjustments, such as the addition of trailing-edge flaps on airfoils to mitigate dynamic-stall pitching moments~\cite{Gerontakos2006dynamic}, and the application of riblets designed to alter boundary layer characteristics for reduced skin-friction drag~\cite{Bliamis2022numerical}. These approaches stem from direct analysis of the fluid dynamics around particular geometric shapes~\cite{Juvinel2023structural}. 
In contrast, airfoil shape optimization addresses the inverse problem by finding specific airfoil geometries to achieve desired fluid behavior~\cite{Juvinel2023structural}.
Continuous advancements in high-performance computing and computational modeling have increased the feasibility of shape optimization using computational fluid dynamics (CFD) simulations, and numerous shape-optimization methods have been designed to improve aerodynamic performance~\cite{He2019robust}. Among these,  gradient descent-based methods---including discrete and continuous adjoint methods~\cite{Dussauge2023reinforcement, jameson1995optimum, nadarajah2001studies, Srinath2010an}---are attractive for their low computational costs in large design spaces. Gradient-free methods excel in finding the global optimum and are well suited for complex optimization tasks involving nonlinear and nonconvex optimization functions~\cite{Skinner2018state, Dussauge2023reinforcement}.  Examples with applications to airfoil shape design include genetic algorithms (GA)~\cite{Chiba2005high, umutlu2022airfoil} and the particle swarm optimization algorithm~\cite{fourie2002particle, nejat2014airfoil}.  

While more complex and often more expensive than passive techniques, active flow control offers significant advantages in achieving higher levels of flow manipulation and is applicable to a broader range of flows. 
One classical technique is the addition of blowing/suction jets, which inject and/or remove air at the airfoil surface to energize the flow and delay separation.
Types include continuous jet actuators (CJAs), synthetic jet actuators (SJAs), and coflow jets (CFJs).
CJAs supply steady, unbroken air streams to/from the boundary layer; many applications may be found in the literature for the National Advisory Committee for Aeronautics (NACA)  0012 airfoil~\cite{Huang2004numerical, Yousefi2015three, Munday2014separation} and NACA 0015 airfoil~\cite{Greenblatt2000the, DEGIORGI2015comparison, SKAROLEK2016700}. These primarily focus on drag reduction for angles of attack $\alpha=5^\circ$ to $\alpha=18^\circ$.
SJAs are zero-mass-flux control devices that alternately inject and remove air through a small slot, with the jet pressure driven by the oscillatory motion of a diaphragm~\cite{Montazer2016Optimization}. This
virtually eliminates the need for plumbing and produces unique effects on base flows that are not possible with steady or pulsed jets~\cite{Smith2003comparison}. Numerical experiments have explored the efficacy of SJAs for NACA 0012~\cite{Wu1998post}, NACA 0015~\cite{DEGIORGI2015comparison, itsariyapinyo2022experimental} and SD 7003 airfoils~\cite{Tousi2021active, TOUSI2022large}. 
Unlike CJAs that are typically positioned at the leading or trailing edge, CFJs are implemented over the majority area of the suction (upper) surface of the airfoil~\cite{Zha2004a}. This is achieved by translating a significant area of the suction surface downward, which substantially enhances lift and reduces drag. Applications on NACA 2415~\cite{Zha2004a}, NACA 0025~\cite{Zha2006effect}, and NACA 0018 airfoils~\cite{vigneswaran2021aerodynamic} have shown significant aerodynamic improvements.
In this study, we adopt a geometrically fixed CJA configuration informed by prior design studies~\cite{Munday2014separation, Yousefi2015three, Huang2004numerical, Tousi2021active}, which allows us to isolate and evaluate the effect of pressure-based control optimization.

Beyond blowing and suction jets, other AFC techniques include plasma actuators and fluidic oscillators.
Plasma actuators utilize ionized gas (plasma) to impart momentum to the surrounding air, thus modifying the boundary layer characteristics and controlling flow separation~\cite{Sundaram2022flow}. For instance,  surface dielectric barrier discharge plasma actuators have been shown to mitigate dynamic stall over periodically pitching NACA 0015 airfoils. This is capable of reducing the degree of flow separation, increasing turbulent transition, and achieving earlier flow reattachment~\cite{yu2020numerical}.
Fluidic oscillators are simple to design and can control separation more effectively than other AFC techniques~\cite{KIM2020Effects}. They can improve lift, reduce drag~\cite{jones2017sweeping}, and diminish downstream flow separation as validated by wind tunnel experiments~\cite{koklu2014flow}.

Although AFC techniques are highly effective, most current approaches rely on iterative evaluation of aerodynamic performance using experimental or numerical testing. Identifying efficient AFC actuators remains a challenge due to two primary factors~\cite{duriez2017machine}.
First, turbulent flows exhibit intricate, nonlinear dynamics and high dimensionality that pose significant challenges for prediction and control.
Second, physical and numerical disturbances, and sensor and actuator noise, challenge the ability of control algorithms to measure and optimize flow conditions~\cite{Wang2022deep, Tadjfar2017active}.
Evolutionary algorithms like GA are adept at solving complex problems and finding global optima~\cite{Tousi2021active}, but these algorithms are often limited by their increasing cost with increasing number of design variables, and they suffer from slow convergence when coupled to CFD simulations~\cite{Dussauge2023reinforcement}.

Recently, machine learning (ML) approaches to flow control have emerged as promising alternatives~\cite{duriez2017machine, Li2022recent}. One approach involves constructing surrogate models for the flow dynamics. For example, Lee \etal~\cite{lee1997application} used a neural network to approximate a mapping between wall shear stress and control jet, then approximated the optimal actuation to achieve 20\,\% skin-friction drag reduction. 
Another prominent ML approach for AFC is deep reinforcement learning (DRL), which is widely used for complex decision-making problems (originally associated with games) and is increasingly used in physical systems~\cite{Rabault2019artificial, garnier2021review,liu2024adjoint}.  For instance, Wang \etal~\cite{Wang2022deep} applied DRL to synthetic jet control over a NACA 0012 airfoil under weak turbulent conditions. 
Sonoda \etal~\cite{sonoda2023reinforcement} and Lee \etal~\cite{lee2023turbulence} reduced turbulent drag in channel flows by varying jet blowing/suction using velocity and shear-stress measurements. Portal-Porras \etal~\cite{PortalPorras2023Active} trained a DRL agent to control flap rotation on a NACA 0012 airfoil. Despite these advancements, DRL faces challenges related to sample inefficiency, difficulty in designing reward functions, and the lack of stability and convergence theories~\cite{li2017deep}.

Another recent advancement in ML for fluid dynamics is the use of solver-embedded optimization---i.e., optimizing over the governing partial differential equations (PDEs)---to achieve model-consistent learning~\cite{Duraisamy2020perspectives}. One version, the deep learning PDE model (DPM) approach~\cite{Sirignano2020DPM}, embeds untrained neural networks in the governing PDEs and trains them using adjoint-based, PDE-constrained optimization. DPM has been successful for turbulence modeling in large-eddy simulation (LES)~\cite{Sirignano2020DPM,MacArt2021Embedded,sirignano2023bluff,hickling2024large} and Reynolds-averaged Navier--Stokes (RANS) simulations~\cite{sirignano2023pde} as well as subcontinuum modeling in hypersonic flows~\cite{nair2023deep, kryger2024optimization}.
In the context of AFC, DPM has been shown to optimize more-effective controllers than DRL and supervised ML for two-dimensional (2D) laminar flows~\cite{liu2024adjoint}. 
DPM's optimization efficiency arises due to its use of adjoint fields to compute the end-to-end sensitivities needed for neural network optimization. The PDE constraint ensures that learned models work alongside  the system's dynamics as modeled by the full-order PDEs. The global optimality convergence of DPM has been established for elliptic PDEs~\cite{sirignano2023pde}, and current research focuses on its effectiveness for nonlinear parabolic and hyperbolic PDEs.
This study presents the first application of adjoint-based active flow control to three-dimensional (3D) turbulent airfoil flow problems. Our approach optimizes a neural network flow controller, with the objective of maximizing the time-averaged lift-to-drag ratio for minimal additional energy, while solving the flow PDEs. 

We first assess the active control framework by optimizing over DNS of two-dimensional (2D) airfoil flows. This provides a controlled, computationally efficient environment for validating the DPM framework. For the Reynolds numbers we consider, the key flow structures---including the laminar separation bubble, shear-layer instabilities, and vortex-shedding dynamics---remain predominantly two-dimensional; hence, these initial 2D simulations  provide critical insight into our subsequent extension of the methodology to LES predictions of fully three-dimensional, turbulent flows.

This paper is organized as follows. Section~\ref{sec:numerics} presents the numerical schemes and implementation of the jet controller. Section~\ref{sec:airfoil_sim}  presents solver validation for 2D and 3D airfoil flows. Section~\ref{sec:DPM} introduces the DPM flow-control framework and the objective function used for optimization. 
In Section~\ref{sec:2D}, we train DPM controllers for 2D airfoil flows at angles of attack $\alpha = \{5^\circ, 10^\circ, 15^\circ\}$ with multiple training windows in order to assess the influence of the time horizon used for controller optimization.
Section~\ref{sec:3D} assesses control performance for 3D airfoil flows including a comparison of 3D-trained controllers and 2D-trained controllers applied to 3D flows. Section~\ref{sec:conclusion} summarizes the findings and offers avenues for future research.
\section{Numerical methods} \label{sec:numerics}

\subsection{Governing equations}

We solve the compressible Navier--Stokes equations in dimensionless, conservative form. The nondimensionalization process is the same as that provided by Liu \& MacArt~\cite{liu2024adjoint} with units of length nondimensionalized by the airfoil chord length, such that the dimensionless $c=1$. All subsequent quantities presented here are dimensionless. Let $\bQ=[\rho,\rho\bu,\rho E]^\top$ be the vector of conserved quantities,  where  $\rho$ is the mass density, $\bu\in\mathbb{R}^d$ is the velocity vector of spatial dimension $d$, 
and $E=e+\bu^\top\bu/2$ is the total energy, where $e =  T/(\gamma\Ma^2)$ is the internal energy of the calorically
perfect gas,  $T$ is the temperature, $\gamma=1.4$ is the ratio of specific heats, and $\Ma=0.4$ is the scaling Mach number. The dimensionless ideal gas equation of state is $p = (\gamma-1)\rho T/\gamma$, where $p$ is the pressure.

The evolution of the conserved quantities is governed by
\begin{equation}
  \pp{\bQ}{t} = \bR(\bQ) = \nabla \cdot [ - \bF_i(\bQ) + \bF_v(\bQ) ],
  \label{eq:NS}
\end{equation}
where $\bR$ is the instantaneous PDE residual, and $\bF_i$ and $\bF_v$ are the inviscid and viscous flux vectors
\begin{align}
 \bF_i(\bQ) = \left[\begin{array}{c} 
    \rho\bu^\top \\
    \rho \bu\bu^\top + \frac{p}{\Ma^2} \bI \\
    \rho\bu^\top \left( E + \frac{ p }{\Ma^2 \rho}\right)
  \end{array}\right] \quad \mathrm{and} \quad
 \bF_v(\bQ) =  \frac{1}{\Rey} \left[\begin{array}{c} 
    0 \\
    \sigma \\
    \bu^\top\sigma - \frac{1}{\Ma^2 \Pr} \mathbf{q}
\end{array}\right], 
\label{eq:fluxes}
\end{align}
where $\bI\in\mathbb{R}^{d\times d}$ is the identity matrix, $\sigma=\mu (\nabla \bu+\nabla\bu^\top - \frac{2}{3}(\nabla \cdot \bu)\bI)$ is the viscous stress tensor, $\mu$ is the dynamic viscosity, $\mathbf{q}=\mu\nabla T$ is the heat flux vector, and $\Rey=5\times 10^4$ and $\Pr=0.7$ are the scaling Reynolds and Prandtl numbers, respectively.

\subsection{Numerical discretization}
The governing equations are solved on generalized curvilinear meshes. The PDEs are semidiscretized in the computational plane using five-point, fourth-order central-difference schemes~\cite{Lele1992} that reduce to first-order, one-sided schemes near domain boundaries. Sixth-order, implicit, low-pass spatial filters are applied every time step to remove spurious oscillations. 
Time is advanced using the fourth-order Runge--Kutta method. No-slip boundary conditions are imposed on the airfoil wall, and far-field boundaries are imposed using absorbing layers~\cite{liu2024adjoint}. The flow solver, \code, is Python-native,  MPI-parallelized, and fully GPU-accelerated. Further details may be found in Liu \& MacArt~\cite{liu2024adjoint}. \code\ is fully differentiable using algorithmic differentiation (AD), which enables efficient construction of the adjoint equations needed for AFC optimization. For LES, we employ the wall adapting local eddy viscosity (WALE) model~\cite{Ducros1998wale} with model constant $C_w=0.325$.

\subsection{Jet boundary conditions} \label{sec:pressure-jet}

The blowing/suction jet actuator is modeled using boundary conditions that regulate the total pressure $p_{0J}$ on the airfoil surface.
The air density is obtained using $\rho_{0J} = \frac{ \gamma p_{0J} }{ (\gamma-1) T_{0J} }$, with a prescribed temperature $T_{0J}=1.083 T_\infty$, where $T_\infty$ is the freestream flow temperature.  
Assuming isentropic expansion, the jet sets Dirichlet boundary conditions for velocity, density, and temperature by matching the pressure between the jet and the flow, $p_J = p$.

When  ${p_{0J} } \geq p_J = p$,
air is injected into the flow at matching static pressure. 
We solve for the jet Mach number from the isentropic relation
\begin{equation}
  \Ma_J = \sqrt{   \left[ \left({\color{black}{ \frac{{p_{0J} }}{p} } } \right)^{  \frac{\gamma-1}{\gamma} } -1 \right] \frac{2}{\gamma-1} },  \label{eq:p0tp}
\end{equation}
where $\Ma_J = \min( \Ma_J, 1.0 )$ is bounded to ensure maximally sonic flow.
Based on the pressure ratio $p_{0J} / p$, the density $\rho_{J}$ and static temperature $T_J$ of the jet are obtained by 
\begin{align}
    \rho_{J} &=  \rho_{0J} \left[ 1 + \frac{\gamma-1}{2} \Ma_J^2 \right] ^{-\frac{1}{\gamma-1}} = \frac{ \gamma p_{0J} }{ (\gamma-1) T_{0J} }  \left({\color{black}{ \frac{{p_{0J} }}{p} } } \right) ^{\frac{1}{\gamma}}  \label{eq:rho_J}, \\
    T_J &= T_{0J} \left( 1+ \frac{\gamma-1}{2}\Ma_J^2 \right)^{-1} 
    = T_{0J} \left({\color{black}{ \frac{{p_{0J} }}{p} } } \right) ^{-\frac{\gamma-1}{\gamma}}.  \label{eq:T_J}
\end{align}
These quantities are obtained in a pointwise manner along the airfoil span for 3D simulations.

We assume that the boundary-normal jet velocity magnitude $U_J$ is maximum at the center of the jet, $U_J^{\max}=U_J(x_c)$, and vanishes at the jet boundary, $U_J(x_c \pm 0.5L) = 0$, where $L$ is the jet width. $U_J(x)$ is assumed to have a parabolic, laminar profile 
\begin{equation} \label{eq:parab_blw}
   U_J (x)  = 4 U_J^{\max} \cdot [ x - (x_c-0.5L) ] \cdot [ -x + (x_c+0.5L) ] / L^2.
\end{equation} 
Substituting Eq.~\ref{eq:T_J} and the sonic velocity $c_s = \sqrt{ (\gamma-1) T_J } / \Ma $ into the definition $U_J =  c_s \Ma_J $, we obtain the jet velocity magnitude 
\begin{equation}
    U_J
    = \frac{\Ma_J}{\Ma} \sqrt{ (\gamma-1)  T_{0J} \left({\color{black}{ \frac{{p_{0J} }}{p} } } \right) ^{-\frac{\gamma-1}{\gamma}} }.  \label{eq:umag}
\end{equation}
The jet velocity vector is then $\bu_J = U_J\bn$, where $\bn$ is the unit outward normal vector from the airfoil surface.
Dirichlet conditions $\bQ_J$ are then applied on the jet wall $\partial^J \Omega$,
\begin{equation}
    \bQ_J = \left[\begin{array}{c} 
    \rho_J \\
    \rho_J \bu_J \\
    \rho_J \left ( e_J+ \frac{1}{2}\bu^\top_J\bu_J \right )
  \end{array}\right],
\end{equation}
where $e_J = {T_J}/({\gamma \Ma^2})$ is the dimensionless jet internal energy.
Following a similar procedure, jet suction is specified when  ${p_{0J} } < p$. Dirichlet conditions for suction are computed by replacing $p_{0J}/p$ by $p/p_{0J}$ in Eqs.~\ref{eq:p0tp}--\ref{eq:umag} and reversing the sign of the normal vector.

\section{Validation of airfoil simulations} \label{sec:airfoil_sim}

We first construct a database of 2D NACA 0012 airfoil flows at $\Rey = 5\times 10^4$ and $\Ma = 0.4$ for $\alpha = [5^\circ, 10^\circ, 15^\circ]$. 
This airfoil geometry is chosen due to its extensive use in DNS calculations in the existing literature~\cite{jones2008direct,hickling2024large,Balakumar2017direct}, especially from ~\cite{jones2008direct} which was conducted on C-type grids. Our calculations mirror those outlined in~\cite{hickling2024large}, which were conducted on two O-type grids with maximum radius $R=20$ and nonuniform, elliptic mesh generation to resolve the airfoil boundary layer.

\begin{figure}
    \centering
\includegraphics[height=5cm]{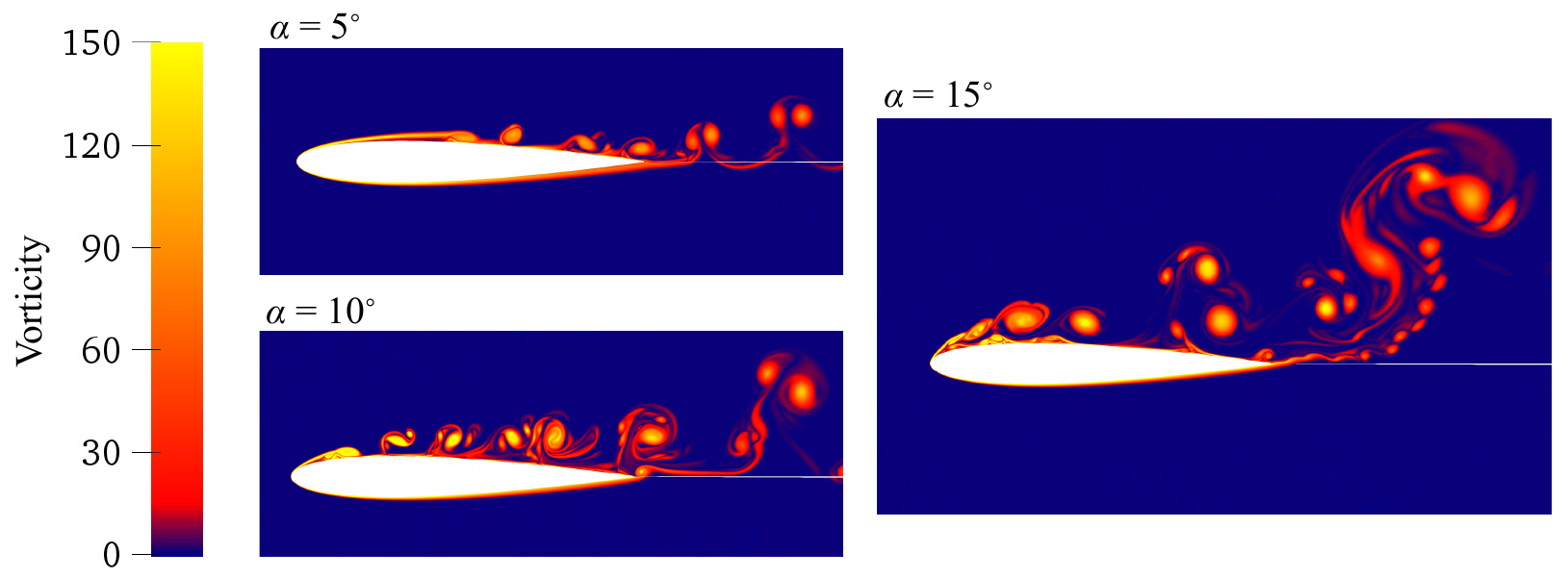}
    \caption{Instantaneous dimensionless vorticity magnitude in uncontrolled 2D airfoil flows  (G1 mesh).}
    \label{fig:afVor}
\end{figure}

Two-dimensional calculations are performed on two grids. The first grid, G1, has $N_\xi=1920$ nonuniform mesh lines along the wall-tangential direction and $N_\eta=540$ nonuniform mesh lines in the wall-normal direction. The second grid, G2, is obtained by downsampling by factors of $2$ in both directions. These mesh sizes, as well as domain extents and maximum resolutions along the airfoil surface,  are listed in Table~\ref{tab:meshgrid}.  Snapshots of the instantaneous 2D velocity magnitude on G1 at three angles of attack are shown in \figref{fig:afVor}. 

Three-dimensional calculations are likewise performed on two grids. DNS calculations use the G1 mesh with $N_z=180$ points in the spanwise direction. For consistency with existing DNS data~\cite{jones2008direct, hickling2024large, Turner2020effect}, all 3D calculations use spanwise domain extent $L_z=0.2$.  LES calculations are performed on a lower-resolution grid, G3, formed by downsampling G1  by factors of 8, 4, and 9 in the tangential, wall-normal, and spanwise directions, respectively, leading to $N_\xi=240$, $N_\eta=130$, and $N_z=20$. The maximum spacings of these meshes along the airfoil surface are listed in Table~\ref{tab:meshgrid}.

\begin{table}
\caption{Simulation parameters: mesh sizes, domain extents, and wall resolutions  of the present calculations compared to those of Jones \etal~\cite{jones2008direct}. All dimensions are given in chord lengths. $R$ is the leading-edge radius and $W$ is the wake length; $W=R$ for O-type grids. $\Delta x^+$, $\Delta y^+$, and $\Delta z^+$ are the maximum mesh spacings in local wall units along the airfoil surface for $\alpha=5^\circ$.}
\centering
\label{tab:meshgrid}
\begin{tabular}{ccccccccccc}
\toprule
                    &    Case                                 & $N_{\xi}$ & $N_{\eta}$ & $N_z$         &  $R$  & $W$   &  $L_z$   & $\Delta x^+$  &  $\Delta y^+$ &  $\Delta z^+$       \\ \midrule
\multirow{3}{*}{2D} & Jones \textit{et al.}~\cite{jones2008direct} & 2570      & 692        &  1   &  7.3  &  5  &  --  & -- & -- &  --            \\ & \textit{PyFlowCL}, G1                     & 1920      & 540        &  1        &  20  &  20  &  --  &  6.19  & 0.683 &  --                \\
                    & \textit{PyFlowCL}, G2                 & 960       & 270        &    1    &  20  &   20 &  -- &  10.1 & 1.14  &  --           \\ \midrule
\multirow{3}{*}{3D}  & Jones \textit{et al.}~\cite{jones2008direct} & 2570      & 692        & \multicolumn{1}{c}{96} &  7.3  &  5  &  0.2   & 3.36   & 1.00  &   6.49 \\
                     & \textit{PyFlowCL}, G1 (DNS)                  & 1920       & 540        & \multicolumn{1}{c}{180}  &  20  &  20  &  0.2   &  4.89  & 0.542  &  3.10  \\
                     & \textit{PyFlowCL}, G3 (LES)                 & 240       & 130        & \multicolumn{1}{c}{20}  &  20  &  20  &  0.2   &  38.0  & 4.76  & 71.7  \\
                    \bottomrule
\end{tabular}
\end{table}

\begin{figure}
  \centering
  \hspace{-0.5cm}
\includegraphics[width=0.45\textwidth]{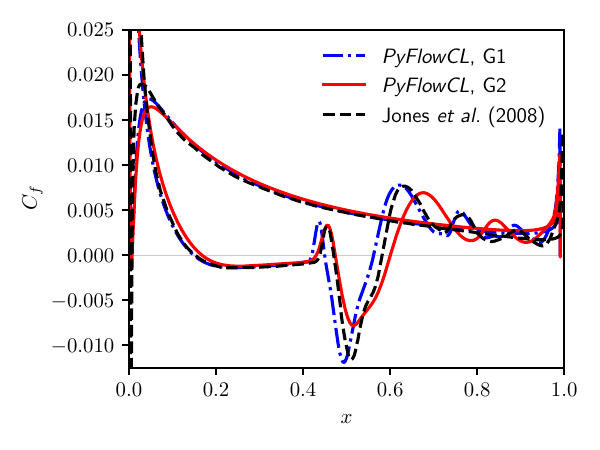}
\includegraphics[width=0.425\textwidth]{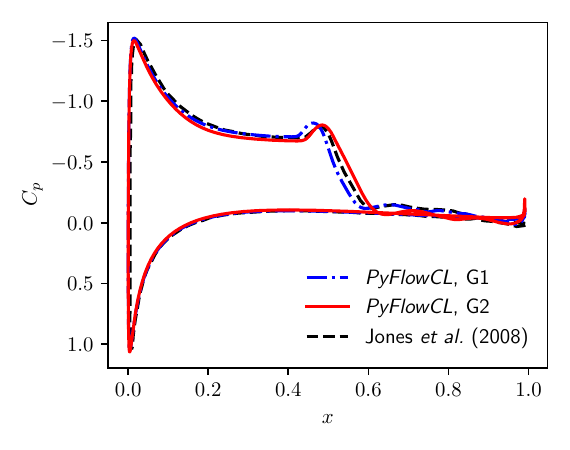}
  \caption{Two-dimensional NACA 0012 flows: Time-averaged friction coefficient $C_f$ (left) and pressure coefficient $C_p$ (right) at $\alpha=5^\circ$ compared to data of Jones \etal~\cite{jones2008direct}.}
  \label{fig:CfCp_2D}
\end{figure}

\figref{fig:CfCp_2D} compares the time-averaged pressure and friction coefficients of the present 2D calculations,
\begin{align}
    C_p&=\frac{p-p_\infty}{\frac{1}{2}\rho_\infty U_\infty^2} &  &\mathrm{and} &
    C_f&=\frac{\tau_w}{\frac{1}{2}\rho_\infty U_\infty^2},
\end{align}
 to those of Jones \etal~\cite{jones2008direct}, where $p_\infty$, $\rho_\infty$, and $U_\infty$  are the freestream static pressure, static density, and velocity of the flow, and $\tau_w$ is the wall shear stress.
The G1 coefficients are well aligned with Jones \etal~\cite{jones2008direct} for $\alpha=5^\circ$, and G2 shows acceptable agreement. Table~\ref{tab:validation} lists separation points $x_\mathrm{sep}$, reattachment points $x_\mathrm{reatt}$, lift and drag coefficients
\begin{align}
    C_l&=\frac{F_L}{\frac{1}{2}\rho_\infty U_\infty^2 A_p} & &\mathrm{and} &
    C_d&=\frac{F_D}{\frac{1}{2}\rho_\infty U_\infty^2 A_p},
\end{align}
where $F_L$ and $F_D$  are lift and drag forces,  and $A_p= c L_z$ is the planform area of the airfoil, 
as well as the time-averaged lift-to-drag ratio $C_l/C_d$ for the three simulations.
Most of these values fall within $\pm 5\,\%$. Since it exhibits adequate performance and reduces costs, we select G2 for subsequent flow controller training and testing.

\begin{table}
\caption{Separation points, reattachment points, lift coefficients, drag coefficients, and lift-to-drag ratios for the present 2D and 3D calculations at $\alpha=5^\circ$ compared to Jones \etal~\cite{jones2008direct}.}
\centering
\label{tab:validation}
\begin{tabular}{ccccccc}
\toprule
                    &    Case                      & $x_\mathrm{sep}$ & $x_\mathrm{reatt}$ & $\overline{C_l}$ & $\overline{C_d}$ & $\overline{C_l/C_d}$ \\ \midrule
\multirow{3}{*}{2D} & Jones \textit{et al.}~\cite{jones2008direct}       & 0.151            & 0.582              & 0.499            & 0.0307        &  16.3  \\
& \textit{PyFlowCL}, G1                                                  & 0.149 (-1.3\%)    & 0.563 (-3.3\%)     & 0.491 (-1.6\%)   & 0.0306 (-3.3\%)  &16.0 (-1.3\%)   \\
                    & \textit{PyFlowCL}, G2                              & 0.164 (8.6\%)   & 0.600  (3.1\%)      & 0.476 (-4.6\%)   & 0.0315 (2.6\%)   &  15.1 (-7.2\%)  \\ \midrule
\multirow{3}{*}{3D} & Jones \textit{et al.}~\cite{jones2008direct}       & 0.099            & 0.607              & 0.621            & 0.0358        & 17.3  \\
       &\textit{PyFlowCL}, G1 (DNS)       & 0.102 (3.0\%)            & 0.595 (2.0\%)               & 0.605 (-0.10\%)          & 0.0356 (-1.5\%)       & 17.2 (-0.8\%)  \\
       &\textit{PyFlowCL}, G3 (LES)       & 0.162 (61\%)            & 0.706 (16\%)               & 0.646 (4.0\%)          & 0.0396 (11\%)       & 16.3 (-5.8\%)  \\ \bottomrule
\end{tabular}
\end{table}

\begin{figure}
  \centering
  \hspace{-0.5cm}
\includegraphics[width=0.45\textwidth]{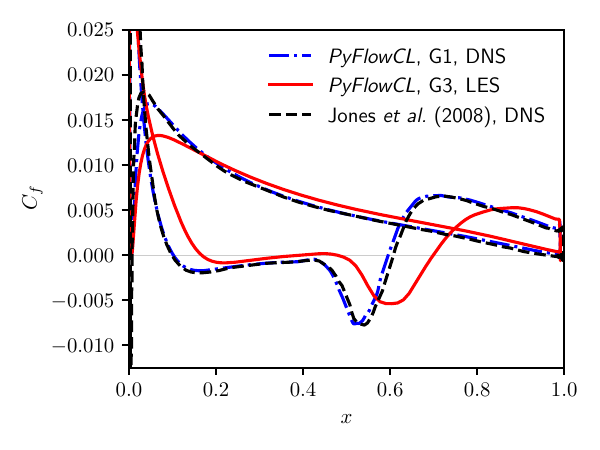}
\includegraphics[width=0.425\textwidth]{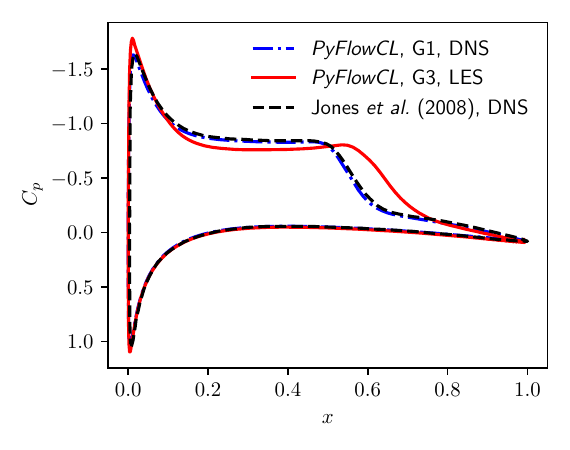}
  \caption{Three-dimensional NACA 0012 flows: Time-averaged G1 (DNS) and G3 (LES) friction coefficients $C_f$ (left) and pressure coefficients $C_p$ (right) at $\alpha=5^\circ$ compared to DNS data of Jones \etal~\cite{jones2008direct}.}
  \label{fig:CfCp_3D}
\end{figure}

The G3 LES is validated by comparing its time-averaged $C_p$ and $C_f$ to those of the G1 DNS and the DNS of Jones \etal~\cite{jones2008direct} in \figref{fig:CfCp_3D}. 
Due to its coarse resolution, the G3 LES  underpredicts the separation and reattachment points; however, it nonetheless reproduces the  lift and drag coefficients, as well as the dominant vortical structures, with acceptable accuracy, within 10\,\% error as reported in Table~\ref{tab:validation}.
Considering the computational cost required by flow controller optimization, we judge the G3 LES to be satisfactory for the purpose of demonstrating the capability of the DPM-based active control strategy. Of course, one could optimize over the G1 DNS at concomitantly higher computational cost.
Additional details of the G1 DNS are available in Hickling \etal~\cite{hickling2024large}, which demonstrates excellent agreement of the present \code\  results with the reference data.

\begin{table}
\caption{Time-averaged lift coefficients $\overline{C_l}$, drag coefficients $\overline{C_d}$, lift-to-drag ratios $\overline{C_l/C_d}$  and jet dynamic pressure $p_{\mathrm{base}}$ for 2D and 3D uncontrolled baseline flows at angles of attack $\alpha=5^\circ,  10^\circ, 15^\circ$.}
\label{tab:base}
\centering
\begin{tabular}{cccccc}
\toprule
       Dimension     & $\alpha$        & $\overline{C_l}$ & $\overline{C_d}$ & $\overline{C_l/C_d}$  & $p_\mathrm{base}$ \\ \midrule
\multirow{3}{*}{2D (G2)} & $5^\circ$  & 0.476           & 0.0315           & 15.1        & 0.632           \\
                    & $10^\circ$ & 0.834           & 0.095           & 8.78         & 0.563           \\
                    & $15^\circ$ & 1.09           & 0.255           & 4.27         & 0.554           \\ \hline
\multirow{3}{*}{3D (G3)} & $5^\circ$  & 0.646           & 0.0396           & 16.3        & 0.632           \\
                    & $10^\circ$ & 0.857           & 0.106              & 8.08         & 0.576           \\
                    & $15^\circ$ & 0.694           & 0.237           & 2.93         & 0.653           \\ \bottomrule
\end{tabular}
\end{table}

Mean aerodynamic performance coefficients of uncontrolled baseline flows at $\alpha = 5^\circ, 10^\circ, 15^\circ$  are listed in Table~\ref{tab:base} as references for control performance analysis.
In the 2D flows, the boundary layer on the upper surface gradually separates from the surface, leading to increased lift and drag, and a decrease of flow pressure in the upper surface. In contrast, for 3D flows, the spanwise component of the flow generates vorticity and spanwise pressure gradients, which impact the pressure distribution and boundary layer behavior. This increases the flow's susceptibility to separation and stall. Consequently, both lift and drag increase at relatively low angles of attack (e.g., $\alpha = 5^\circ$ and $10^\circ$) before reaching stall conditions. However, the flow separation of a higher $\alpha$ becomes excessive, resulting in a sudden decrease in lift and a sharp increase in drag, signaling stall. For $\alpha=15^\circ$, the 3D flow is in a post-stall state in which the flow may partially reattach, resulting in a decrease in lift, increase in drag, and an increase in pressure.

All of the above-averaged data are computed over a simulation interval of 20 chord flow times with 100 and 125 data snapshots per chord flow time for 2D and 3D airfoils, respectively.

\section{Adjoint-based control framework} \label{sec:DPM}

We aim to solve the optimization problem
\begin{equation}\label{eq:obj}
  \min \Bar{J}(\bQ; \theta) \mathrm{\ subject\ to\  } R(\bQ, \Dot{\bQ}, t; \theta) = 0,
\end{equation}
where $\Bar{J}(\bQ,\theta)=\int_{\tau} J(\bQ,t;\theta)dt$ is a time-integrated objective function,  $\tau$ is the optimization window duration, $R(\bQ,\dot \bQ, t; \theta)$ is the instantaneous residual of the governing PDEs,
$\Dot{\bQ} = \partial \bQ/\partial t$, and $\theta\in\mathbb{R}^{N_\theta}$ are
\makeJG{the trainable parameters of the flow controller. We compare two types of control: constant control, in which the control variable is $\theta$ itself, and active control, in which the control variable is obtained by evaluating a neural network with parameters $\theta$. Specific definitions for these for the present control application are provided in Section~\ref{sec:DPM_control}.}

The gradient $\nabla_{\theta} \Bar{J}$ is needed to minimize Eq.~\ref{eq:obj} using stochastic gradient descent; in the DPM approach these gradients are computed using adjoint variables $\hat \bQ(t)$,
    \begin{equation} \label{eq:gradL}
            \nabla_{\theta} \Bar{J} = 
            \int_{\tau} \left( \Hat{\bQ}^\top \pp{R}{\theta} + \pp{J}{\theta} \right) dt,
    \end{equation}
in which $\hat \bQ(t)$ satisfies the linear ordinary differential equation
    \begin{equation} \label{eq:uhat}
         {\color{black}\frac{d\Hat{\bQ}^\top}{dt}}  
         =   \Hat{\bQ}^\top \pp{R}{\bQ} 
         + \pp{J}{\bQ}.
    \end{equation}
This adjoint equation is solved backward in time from $t=\tau$ to $t=0$ with initial condition $\hat \bQ(\tau)=0$, after which the neural network weights may be optimized using the gradients calculated using Eq.~\ref{eq:gradL}. This process is then performed over the next optimization window, $t\in [\tau, 2\tau]$, and is repeated for subsequent time windows until $\bar J$ converges.

In general, the Jacobian $\partial R/\partial \bQ$ in Eq.~\ref{eq:uhat}  is cumbersome to derive analytically and must be rederived upon changes to the governing PDEs and/or physical models (e.g., equations of state). \code\ addresses  these potential sources of error using automatic differentiation over $R(\bQ,\dot \bQ, t; \theta)$ via the \textit{PyTorch} library~\cite{NEURIPS2019_9015}. This leverages computational graph representations to compute the discrete-exact chain-rule derivatives of function outputs with respect to inputs.

Furthermore, the adjoint method is particularly useful in that it enables optimization over arbitrary time horizons $\tau$ without requiring the construction of the computational graph over the entire $\tau$. Since the time horizon to observe meaningful flow changes (as is needed to optimize flow controllers) could span many hundreds or thousands of simulation time steps, constructing the computational graph over the entire $\tau$ could require prohibitively large amounts of computer memory. Instead, the adjoint-based optimization approach only requires the construction of the computational graph over single time steps to advance Eq.~\ref{eq:uhat}. The memory requirement of adjoint-based optimization therefore does not depend on the choice of $\tau$. Further discussion of the computational efficiency of the method may be found in Liu \& MacArt~\cite{liu2024adjoint}. We assess the influence of $\tau$ for 2D airfoil flow controller training in Section~\ref{sec:2D_tau_analysis}.
\makeFG{The adjoint-based optimization framework for active flow control was developed by Liu and MacArt~\cite{liu2024adjoint}; the novel contribution of this work is its application to turbulent 3D flows.}

\subsection{Airfoil flow control framework} \label{sec:DPM_control}

The control objective is to minimize the time-averaged drag-to-lift ratio (equivalently, to maximize the lift-to-drag ratio) by applying boundary-layer injection/suction at the upper airfoil surface. 
We construct a continuous objective (loss) function
\begin{equation}
    \Bar{J} (\bQ;\theta) = \frac{1}{\tau} \int_{\tau} J(\bQ; \theta) dt = \frac{1}{\tau} \int_{\tau} \beta_1 (C_d/C_l)^2 + \beta_2\sinh^2(\Ma/a)  dt,   \label{eq:loss}
\end{equation}
in which $J$ is the instantaneous loss, $\tau=100 \Delta t = 0.002$ is the optimization time horizon with $\Delta t=2\times10^{-5}$ the simulation time step size, and the hyperparameters are $\beta_1=2.0\times10^8$, $\beta_2=200$, and  $a=0.17$. These are chosen to give $J\times\Delta t = O(1)$. The last term in Eq.~\ref{eq:loss} is a penalty to minimize large control energy expenditures by confining the jet Mach number with a hyperbolic sine function $\sinh(x)$.

\begin{figure}
 \centering
\includegraphics[width=1.0\linewidth]{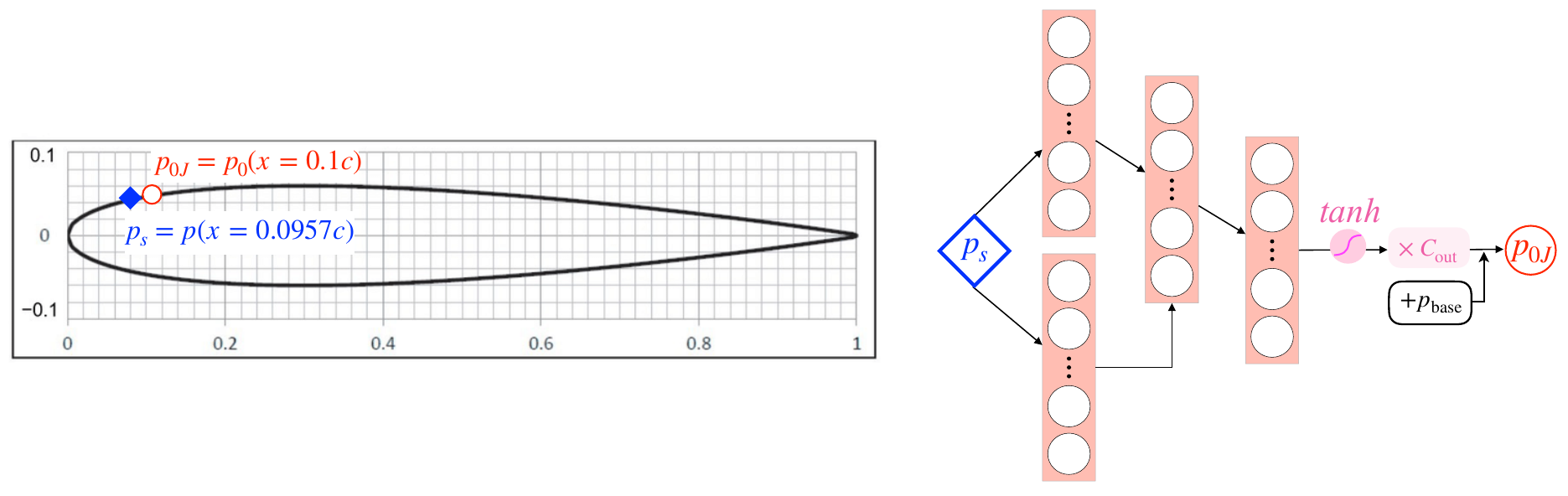}
\caption{Illustration of the DPM-based active-control framework. Left: location of pressure sensor ($p_s$) and active jet controller ($p_{0J}$) along the airfoil suction surface. Right: feedforward neural network structure with a factor $C_\mathrm{out}=0.5$ on the hyperbolic-tangent output layer.}\label{fig:af_NN}
\end{figure}

We implement the control by manipulating the total pressure $p_{0J}$ of an assumed gas reservoir. Depending on the magnitude of the controlled pressure and the boundary-layer static pressure, air is injected into the domain for  ${p_{0J} } \geq p$ or removed from the domain for  ${p_{0J} } < p$. 
The isentropic jet model is provided in Sec.~\ref{sec:pressure-jet}.
The controlled jet total pressure pressure is obtained as the sum of the time-averaged uncontrolled dynamic pressure $p_{\mathrm{base}}$ (reported in Table~\ref{tab:base}) and a neural network pressure modification $F_{\theta}(p_s,\theta)$  with a single, scalar input $p_s$:
\begin{align}
     p_{0J} = p_{\mathrm{base}} + F_{\theta}(p_s,\theta). \label{eq:af_p} 
\end{align}
The feed-forward neural network, shown in \figref{fig:af_NN}, comprises a four-layer, fully connected, dense network with $H=200$ hidden units per layer. Activation functions are specified as rectified linear units (ReLU) for hidden layers and hyperbolic-tangent ($\tanh$) functions with a factor $C_\mathrm{out}=0.5$ for the output layer.
For 2D airfoils, $p_\mathrm{base}$ decreases gradually due to gradual flow separation from $5^\circ$ to $15^\circ$.
For 3D airfoils, $p_\mathrm{base}$ increases for post-stall conditions ($\alpha\gtrsim 15^\circ$) at which the flow may partially reattach.

The input to the neural network is the sensed static flow pressure $p_s$ at $x_s=0.0975$ on the suction (upper) surface. 
The control jet ($p_{0J}$) has width $ L = 0.02$ and is located at $x_c=0.1$, which is close to the natural separation point for 3D flows. These geometry parameters are selected following studies of jet location and size~\cite{Munday2014separation, Yousefi2015three, Huang2004numerical, Tousi2021active} to achieve effective control performance. For 2D flows, the jet is a slot along the semi-infinite upper surface. For 3D flows, the jet expands to fill the spanwise direction (``line controller''). 
Details of these are given in Section~\ref{sec:PCLC}.

\makeFG{
The neural network parameters are updated using the RMSprop optimizer~\cite{RMSProp} with a learning rate $\mathrm{LR}_{k}$ at the $k^\mathrm{th}$ step:
\begin{equation}
    \theta_{k+1} = \theta_k - \mathrm{LR}_{k} \frac{\nabla_{\theta} \Bar{J}}{ \sqrt{E_k + \epsilon} },
\end{equation}
in which $E_k = \chi E_{k-1} + (1-\chi) (\nabla_{\theta} \Bar{J})^2$ is the exponentially weighted average of the squared gradients with the initial $E_{0}=0$, $\chi=0.99$ is the exponential decay rate, and $\epsilon=1\times 10^{-8}$ is added to the denominator to improve numerical stability.
The learning rate is initialized at $\mathrm{LR}_0=5\times10^{-5}$ and decays by $\Delta_\mathrm{LR}$ every $s$ iterations utilizing the scheduler  $\mathrm{LR}_{k} = \mathrm{LR}_0 \times \Delta_\mathrm{LR}^{ \lfloor k/s\rfloor}$.}
\makeFG{The RMSprop optimizer is chosen due to its adaptive learning rate adjustment using the gradient magnitude history, which performs well for nonconvex optimization.}

Training is parallelized using message passing interface (MPI) domain decomposition in the wall-normal direction. The 2D cases (G2) require approximately 13 wall-clock hours per chord-flow time on 16 MPI ranks (AMD EPYC 7532), and the 3D cases (G3) require approximately 14 wall-clock hours per chord-flow time on 32 MPI ranks.

\subsection{Power coefficient and corrected aerodynamic efficiency} \label{sec:power}
In the AFC system, energy consumption is considered through a dimensional jet power consumption coefficient $\hat{P}_J$, which has the dimensionless form $P_J$ by normalizing with the freestream flow power
\begin{align}
  P_J = \frac{\hat{P}_J}{\frac{1}{2}\rho_{\infty} u_{\infty}^3 A_p}.
\end{align}
The dimensionless jet power consumption coefficient $P_J$ is obtained through the mass flow rate of the jet $\Dot{m}_J$ and the specific enthalpy change $\Delta H$
\begin{align}
  P_J= \Dot{m}_J  \Delta H = \rho_J |\bu_J| A_J \Delta H,
  \label{eq:control_power}
\end{align}
where $A_J = L \times z_J $ is the jet area with $z_J=1.0$ for 2D airfoils,
and $z_J=L_z=0.2$ for 3D line controllers.
For a calorically perfect gas,
\begin{equation}
    \Delta H = \Delta h + \Delta \left(\frac{1}{2} |\bu_J|^2 \right) = c_p \Delta T + \Delta \left(\frac{1}{2} |\bu_J|^2 \right),
\end{equation}
where $\Delta h$ is the internal energy change and $c_p$ is the heat capacity.
Typically, electric fans supply this power with an efficiency $\eta_p = 40\,\%$ for relatively efficient synthetic jets~\cite{li2011energy, Girfoglio2015Modelling}. We also assume in this study to represent the required power
\begin{align}
  P_c = P_J/\eta_p .
\end{align}

For a conventional airfoil, the wing aerodynamic efficiency is defined as $C_l/C_d$. However, the efficiency is modified for AFC to consider the energy consumption. Following Barrios \etal~\cite{Barrios2022Simulation}, the corrected aerodynamic efficiency is
\begin{equation}
    \left( \frac{C_l}{C_d} \right)_c =  \frac{C_l}{ (C_d)_c}  = \frac{C_l}{C_d + P_c},
\end{equation}
where $(C_d)_c$ is the equivalent drag coefficient that includes the drag of the aircraft system and the power required by the control jet.

\section{Control performance for 2D airfoils} \label{sec:2D}

We now develop DPM-based flow controllers to improve the lift-to-drag performance of 2D airfoils at $\alpha=5^\circ$, $10^\circ$, and $15^\circ$. 
These 2D cases are physically meaningful at the Reynolds numbers considered, for which the laminar separation dynamics remain predominantly two-dimensional even for fully three-dimensional flows. The 2D setting enables high-fidelity DNS-based training and provides a well-controlled environment for exploring the behavior of the DPM approach before extending to 3D turbulent flows in Section~\ref{sec:3D}.
We first investigate the control effectiveness of the learned controllers in Section~\ref{sec:2D_per} by comparing lift, drag, and lift-to-drag coefficients with the uncontrolled baseline flows for in-sample $\alpha$ but for out-of-sample test time windows.
We then assess the influence of optimization window size in Section~\ref{sec:2D_tau_analysis} for $\alpha=5^\circ$. We find that optimizing over short training intervals (25 to 200 time steps) is able to train effective adaptive controllers, but going to longer time windows results in slower-converging controller parameters and progressively less effective controllers.

\subsection{Control effectiveness} \label{sec:2D_per}

The DPM controller is trained for one chord-flow time  ($t_\mathrm{train}=1$) with simulation time step size $\Delta t = 2\times 10^{-5}$, optimization window size $\tau = 100\Delta t$, and $N_\mathrm{itr}=500$ total optimization iterations. 
After training separate controllers for the three values of $\alpha_\mathrm{train}$, we perform \emph{a posteriori} tests for in-sample angles of attack ($\alpha_\mathrm{train}=\alpha_\mathrm{test}$).
\makeFG{Even for in-sample angles of attack, all \emph{a posteriori} tests are out-of-sample in time  ($t_\mathrm{test}=16$), that is, they have much longer duration than any individual optimization window.}
In general, the learned controller creates a suction jet which increases lift, decreases  drag, and causes significant qualitative flow changes.

\subsubsection{Qualitative changes to the controlled flow} \label{sec:trate2D_flow}

\begin{figure}
\centering
\includegraphics[width=1.0\linewidth]{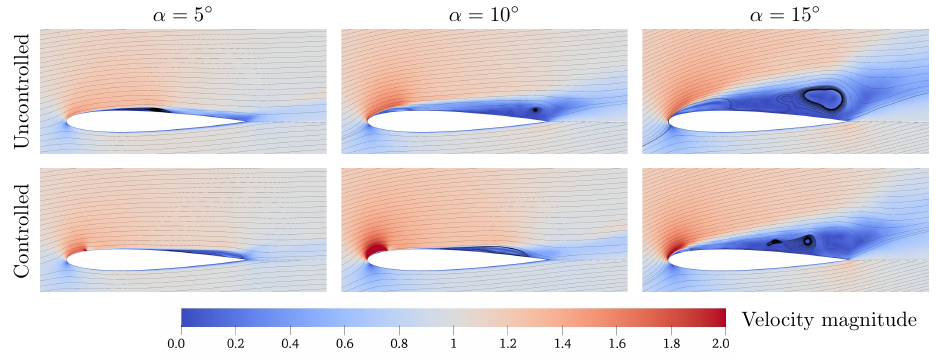}
\caption{Velocity magnitudes and streamlines of time-averaged uncontrolled and controlled 2D airfoil flows.}\label{fig:2D_streamline}
\end{figure}

\figref{fig:2D_streamline} illustrates the qualitative changes to the averaged velocity magnitude along with streamlines in uncontrolled and in-sample controlled flows.  The streamlines of the controlled flows illustrate smaller and delayed separation bubbles compared to the baseline cases. 
For $\alpha=5^\circ$, the controlled separation bubble moves aft to $0.5<x<0.8$. For $\alpha=10^\circ$, the leading-edge separation bubble is diminished and is pushed to an aftward, elongated shape. For $\alpha=15^\circ$, the large laminar separation bubble breaks into several smaller bubbles.
In this latter case, the suction jet reduces the normal momentum and weakens the trailing-edge vortex strength, which accelerates the leading-edge flow  and causes it to remain better attached to the upper surface. Similar results can be found in  suction control of an airfoil at $\alpha=16^\circ$~\cite{Fatahian2019comparative}, $\alpha=18^\circ$~\cite{Huang2004numerical}, and in wind turbine blades at low angles of attack~\cite{Chen2023A}.

\begin{figure}
 \centering
\includegraphics[width=1.0\linewidth, trim=0 15 0 0, clip]{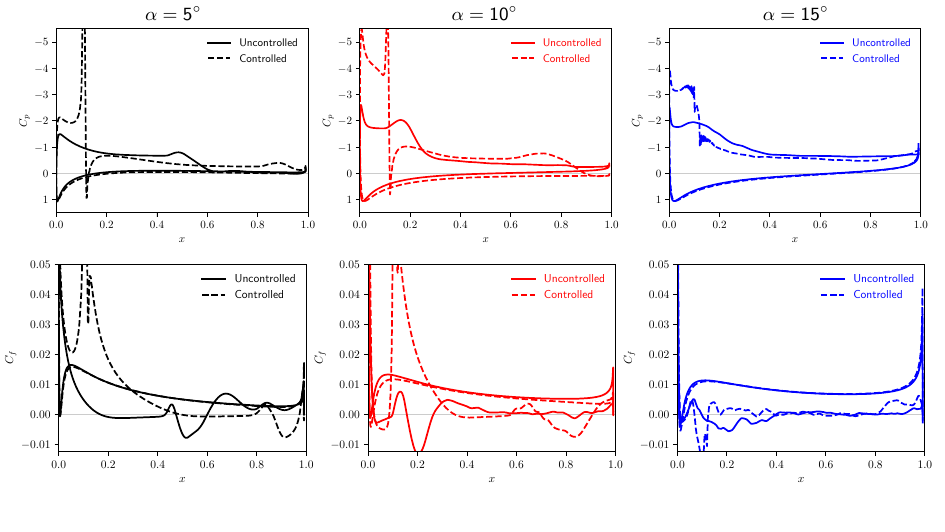}
\caption{Pressure and friction coefficients for uncontrolled and controlled 2D airfoil flows.}\label{fig:2D_Cp_Cf_cmp}
\end{figure}

As shown in  \figref{fig:2D_Cp_Cf_cmp}, the controlled $C_p$ exhibits a spike around the jet, which mitigates the peak negative pressure zone on the upper front edge and amplifies the pressure difference between the upper and lower surfaces. This leads to an increase in lift. Notably, the peak negative pressure is reduced from -1.5 to -2.2 at $\alpha=5^\circ$, -2.5 to -5.5 at $\alpha=10^\circ$, and -2.6 to -4.0 at $\alpha=15^\circ$, respectively. The $\alpha=10^\circ$ flow exhibits the largest change, corresponding to the most significant aerodynamic performance improvement.
The $C_p$ distribution undergoes considerable changes near the leading-edge area, with an adjustable downstream flow. A higher spike in the jet area results in lower downstream pressure compared to the uncontrolled flows, which is particularly evident for $\alpha=5^\circ$ and $\alpha=10^\circ$. However, the influence of the enlarged negative pressure zone at $\alpha=15^\circ$ is relatively localized, with minimal changes in downstream flow. We hypothesize that a higher suction power would be more effective at this high angle of attack~\cite{Fatahian2019comparative, Chen2023A}. 
These findings align well with the results of Huang \etal~\cite{Huang2004numerical}, who demonstrated that a suction jet placed near the leading edge at $x_c=0.1c$ is most effective in manipulating the boundary layer to increase lift.

\subsubsection{Drag and lift} \label{sec:trate2D_ClCd}

\begin{figure}
\centering
\includegraphics[width=1.0\linewidth, trim=11 10 0 2, clip]{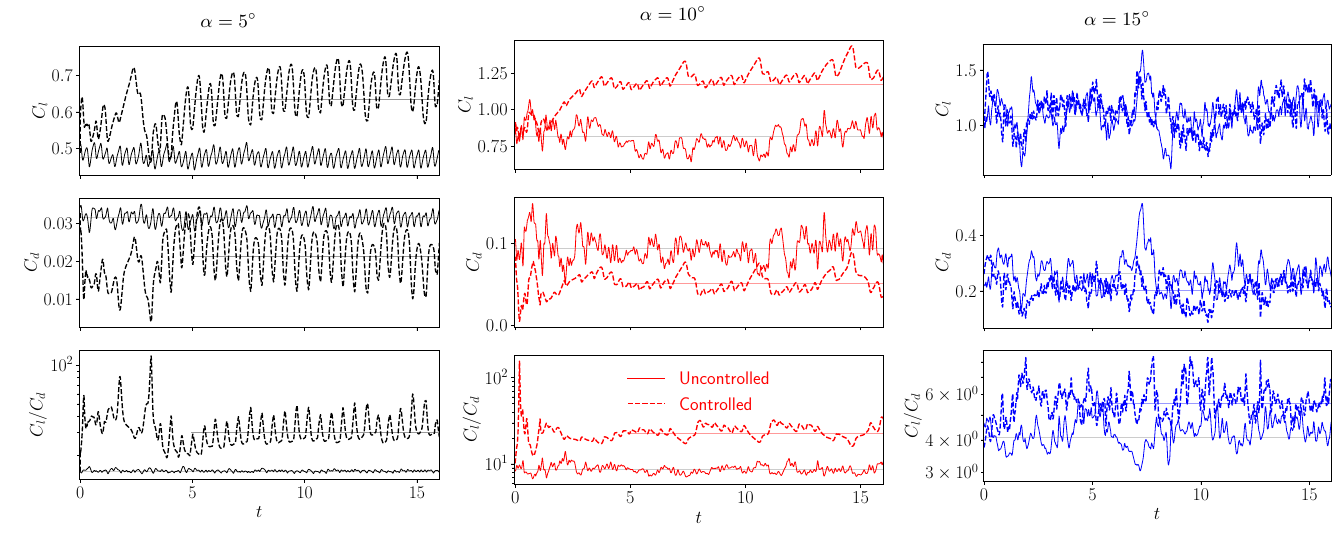}
\caption{Instantaneous and time-averaged lift coefficients, drag coefficients, and lift-to-drag ratios for 2D uncontrolled (solid lines) and controlled (dashed lines) airfoil flows. Time averages are computed for $5<t\leq16$.}\label{fig:ClCdr}
\end{figure}

\figref{fig:ClCdr} shows the time history of the instantaneous uncontrolled and adaptively controlled (DPM) lift and drag coefficients and the instantaneous lift-to-drag ratio. The time-averaged values, $\overline{C_l}$, $\overline{C_d}$, and $\overline{C_l/C_d}$, are tabulated in Table~\ref{tab:2D_mClCd}. The effectiveness of the DPM online controller is demonstrated by its significant lift increase and drag reduction for all angles of attack. Notably, the DPM-trained controllers significantly reduce instantaneous drag and improve instantaneous lift within a span of $0.5$ time units. After an initial transition period ($t\leq 5$), the controlled wake stabilizes, which for $\alpha = 5^\circ$ and $10^\circ$ is periodic. While the controlled velocity fluctuations decrease for $\alpha = 15^\circ$ compared to the uncontrolled flow, the controlled flow remains chaotic.

In comparison to the uncontrolled baseline, the controlled flow exhibits lift-coefficient increases of approximately 40\,\% to 50\,\% and drag-coefficient reductions of 30\,\% to 45\,\% for flows between $\alpha = 5^\circ$ and $10^\circ$. These improvements result in a two- to threefold improvement in $\overline{C_l/C_d}$. These significant performance improvements are comparable to the suction performance reported by Huang \etal~\cite{Huang2004numerical}. However, the control energy consumption at $\alpha = 5^\circ$ is relatively high, leading to only marginal improvements in corrected aerodynamic performance $\overline{(C_l/C_d)_c}$. For the flow at $\alpha = 10^\circ$, the control energy is lower, which leads to a substantial improvement in $\overline{(C_l/C_d)_c}$. At $\alpha = 15^\circ$, the controller shows minimal effect on lift but achieves a noticeable drag reduction. The energy consumption remains low and results in improvements in both $\overline{C_l/C_d}$ and $\overline{(C_l/C_d)_c}$.

\begin{table}
\caption{Control performance of \makeFG{adaptive} and constant DPM controllers for 2D airfoil flow: Time-averaged lift coefficient, drag coefficient, lift-to-drag ratio and corrected lift-to-drag ratio (\S\ref{sec:power}). “Mag.” indicates magnitude; “Inc. (\%)”  indicates the percent increment of controlled flows versus the uncontrolled baseline.}
\label{tab:2D_mClCd}
\centering
\begin{tabular}{cccrcrcrcr}
\toprule
\multirow{2}{*}{$\alpha$}   & \multirow{2}{*}{Controller type}  & \multicolumn{2}{c}{$\overline{C_l}$} & \multicolumn{2}{c}{$\overline{C_d}$} & \multicolumn{2}{c}{$\overline{C_l/C_d}$} & \multicolumn{2}{c}{$\overline{(C_l/C_d)_c}$} \\ \cline{3-10} 
                            &                                  & Mag.     & \multicolumn{1}{c}{Inc.}  & Mag.     & \multicolumn{1}{c}{Inc.}  & Mag.       & \multicolumn{1}{c}{Inc.}    & Mag.         & \multicolumn{1}{c}{Inc.}      \\ \midrule
\multirow{2}{*}{$5^\circ$}  & \makeFG{Adaptive}                            & 0.660   & 38\%                    & 0.0225   & -29\%                   & 29.4    & 95\%                      & 15.5      & 2.8\%                        \\
                            & Constant                          & 0.661   & 39\%                    & 0.0225   & -29\%                   & 29.4    & 95\%                      & 15.6      & 3.1\%                        \\
                  &  \makeFG{ Constant-Opt}                         &  \makeFG{0.648}   &   \makeFG{ 36\%}            & \makeFG{0.0218}   &  \makeFG{-31\%}           
                               & \makeFG{29.7}            &  \makeFG{ 97\%}    &  \makeFG{12.2}     &  \makeFG{-19\%}  \\ \midrule
\multirow{2}{*}{$10^\circ$} & \makeFG{Adaptive}                            & 1.24   & 51\%                    & 0.0944   & -44\%                   & 23.5    & 171\%                     & 20.1     & 132\%                       \\
                            & Constant                         & 1.25   & 52\%                    & 0.0555   & -41\%                   & 22.4    & 159\%                     & 19.4      & 125\%                       \\ \midrule
\multirow{2}{*}{$15^\circ$} & \makeFG{Adaptive}                           & 1.11   & 2.9\%                     & 0.199   & -25\%                   & 5.60     & 36\%                      & 5.52       & 34\%                        \\
                            & Constant                         & 1.10   & 1.2\%                     & 0.195   & -26\%                   & 5.63     & 37\%                      & 5.58       & 36\%                        \\ \bottomrule
\end{tabular}
\end{table}

\subsection{Assessment of constant-pressure controllers}

For the present 2D cases, the minimal time variation of the adaptive controller suggests the use of a constant-pressure controller. We  consider two types of constant controllers. 
\begin{enumerate}[label={(\roman*)}]
    \item ``\textit{Constant-Avg}'': this controller, denoted subsequently simply as  ``\textit{Constant},'' is obtained by time-averaging the adaptive, neural network-controlled total pressure over $5 < t \leq 16$:   $p_{0J}^\mathrm{const} = \langle p_{0J}(p_s, p_\mathrm{base};\theta) \rangle_{t=5\sim 16}$.
    
    \item  ``\textit{Constant-Opt}'': this controller is a DPM-optimized, constant-parameter controller,  trained for one chord-flow time ($t_\mathrm{train} = 1$) at $\alpha = 5^\circ$ with the same objective function, optimizer, and learning rate as the adaptive neural network model described in Section~\ref{sec:DPM_control}. 

\end{enumerate}
The performance of the two simplified controllers, which do not rely on the sensed pressure, is included in Table~\ref{tab:2D_mClCd}. The performance of the \emph{Constant-Avg} controllers is comparable to that of the \makeFG{adaptive} controllers for these 2D flows, which suggests that the flow at the leading edge remains relatively stable. The {\textit{Constant-Opt}} controller has similar influence on drag and lift to the \textit{Constant-Avg} and adaptive controllers; however, the energy consumption of this optimized single-parameter controller is relatively high, resulting in a $19\,\%$ reduction of $\overline{(C_l/C_d)_c}$. This comparison indicates that the adaptive controller learns different, more efficient control dynamics than the optimized single-parameter controller, and that the time-averaged output of the adaptive controller sufficiently approximates this control for 2D cases. Conversely, for 3D flows, the performance of the adaptive controllers is comparatively better than that of the constant-pressure controllers (Section~\ref{sec:2Dtravs3Dtra}).

\subsection{Influence of optimization window size}  \label{sec:2D_tau_analysis}

The choice of the DPM training window $\tau$ can significantly affect training convergence and control performance. We assess its influence by training DPM controllers for the $\alpha=5^\circ$ flow over $ \tau/\Delta t \in [ 25, 6400]$ time steps ($\tau \in [ 5\times10^{-4}, 1.28\times10^{-1}]$ chord flow times).

\begin{figure}
\centering
\includegraphics[width=0.7\linewidth, trim=0 15 0 10, clip]{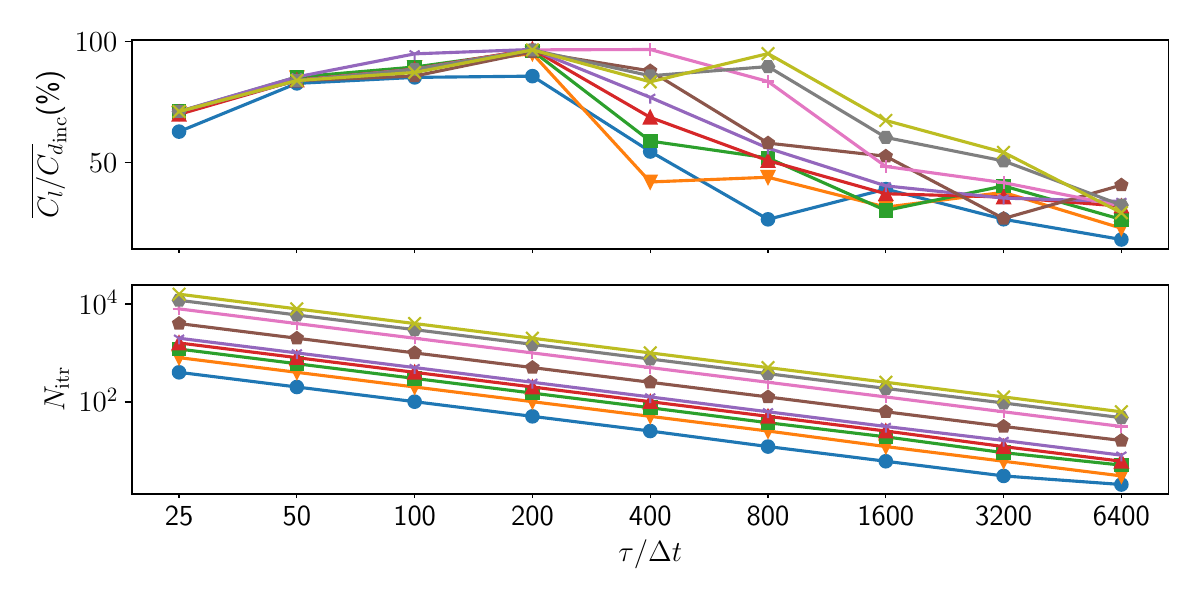}
\includegraphics[width=0.14\linewidth, trim=725 15 0 10, clip]{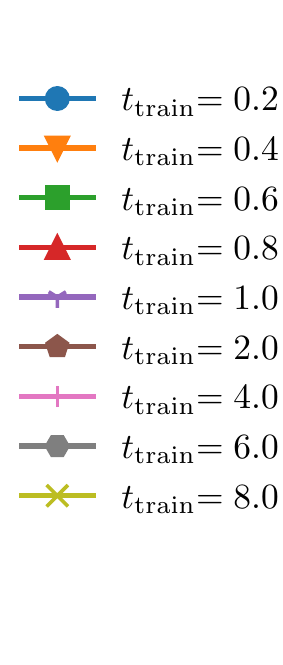}
\caption{Percentage increment of controlled flow performance and required number of optimization iterations $N_\mathrm{itr}$ for different optimization window sizes $\tau$ and training durations $t_\mathrm{train}$. Each value is obtained by averaging results from two randomly initialized neural network models.}\label{fig:tau}
\end{figure}

\figref{fig:tau} displays the time-averaged lift-to-drag percentage increment $\overline{C_l/C_d}_\mathrm{inc}$ versus $\tau$ in \emph{a posteriori} tests; two randomly initialized neural network models were optimized and averaged to obtain the mean values. Each line represents the same training duration $t_\mathrm{train}=\tau N_\mathrm{itr}$, where larger $\tau$ corresponds to fewer optimization iterations $N_\mathrm{itr}$. Thus, the active-control models converge faster for the smaller values $\tau/\Delta t \in [50, 200]$ ($\tau \in [ 1\times10^{-3}, 4\times10^{-3} ]$), resulting in the best control performance. In these cases, stopping training early can yield excellent results with reasonable computational cost. 
However, when $\tau/\Delta t=25$ ( $\tau = 5\times 10^{-4}$), the optimization window is too short to fully observe the system’s dynamics.

Although control performance improves with more iterations for intermediate windows $ 200 < \tau/\Delta t \leq 1600$ ($4\times10^{-3} <\tau\leq 3.2\times10^{-2} $), achieving convergence is challenging for these windows, 
as the adjoint method struggles to provide accurate gradients for long-time-averaged objective functions in chaotic systems~\cite{talnikar2016unsteady}.
This is becomes more apparent for cases the longest windows $\tau/\Delta t=3200$  and $\tau/\Delta t=640$ ($\tau=2.56\times 10^{-1}$ and $1.28\times 10^{-1}$), for which the learned controllers are the least effective. 
When retraining a  $\tau/\Delta t=200$-optimized controller for $\tau/\Delta t=6400$, the controller efficacy decreases by 2\,\% and 13\,\% for $t_\mathrm{train}=2$ and $4$, respectively, which further illustrates the difficulty of optimizing over long time horizons of chaotic flow using the adjoint method. 
These findings are consistent with those of Liu \& MacArt~\cite{liu2024adjoint}, who found small optimization windows to be effective for nonlinear dynamics and the attendant adjoint calculations.

\section{Control performance for 3D airfoils} \label{sec:3D}

We now evaluate the performance of sensor-based controllers across a range of angles of attack in 3D turbulent flows. We introduce a line controller with spanwise-local actuation and compare the performance of controllers trained on 2D flows (Section~\ref{sec:2D}) with that of adaptive and constant-pressure controllers trained directly on 3D flows. The 3D-trained adaptive controller is the only controller tested that consistently improve on $\clcd$ and $\clcdc$ (compared to the uncontrolled flow) across the full range of angles of attack tested. In particular, for the fully separated flow at $\alpha = 15^\circ$, it provides 50\,\% greater improvement of $\clcdc$ compared to its constant-pressure counterpart, illustrating the effectiveness of adaptive controllers enabled by the training methodology.

\subsection{Line controller configuration} \label{sec:PCLC}

The line controller imposes spanwise-local jet control along the spanwise direction $-0.1c \leq z_j\leq 0.1c$ at streamwise location $x_j=0.1c$. This design mimics the distributed suction and blowing techniques commonly used in aerodynamic flow control. Neural network inputs for the DPM controller are spanwise-local pressures at $x_c = 0.0975$, i.e., upstream of the control jet, with the DPM controller learning an optimum control distribution along the spanwise direction.

Actuator settings used in \emph{a posteriori} 3D airfoil tests are listed in Table~\ref{tab:3Dcase}. The controller type (``LC'') suffixed by a ``2'' indicates a controller trained for 2D flows, while a suffix ``3'' indicates a controller trained for 3D flows. Tests of 2D-trained controllers are out-of-sample for flow dimensionality. The 2D- and 3D-trained models are each evaluated pointwise along the spanwise direction, with full spanwise variation in their controlled jet total pressures permitted.

The time-averaged, constant controller $p_{0J}^\mathrm{const} = \langle p_{0J}(p_s, p_\mathrm{base};\theta) \rangle_\mathrm{t=10\sim 30}$ is constructed by averaging the online, DPM-controlled total pressure $p_{0J}$ over $t=10 \sim 30$, skipping the initial transition stage $t<10$. These simplified, constant actuators do not rely on any neural network inputs and are denoted by an additional suffix ``c.''
All of the controllers are tested for 3D flows for test duration $t_\mathrm{test}=30$.

\begin{table}
\caption{Training and testing cases for 3D NACA 0012 flows. ``LC'' indicates a line controller; ``2'' indicates 2D training; ``3'' indicates 3D training; and ``c'' indicates time-averaged (constant-pressure) control actuation. All controllers are tested for 3D flows.
}
\centering
\label{tab:3Dcase}
\begin{tabular}{lcc}
\toprule
Case & Controller Type & Training Dimension                        \\ \midrule
UC   & Uncontrolled      & -                                               \\
LC2  & Line -- \makeFG{Adaptive}            & 2D                                  \\
LC2c & Line -- Constant           & -                                             \\
LC3  & Line -- \makeFG{Adaptive}            & 3D                                  \\
LC3c & Line -- Constant           & -                                              \\ \bottomrule
\end{tabular}
\end{table}

\subsection{Extended performance of 2D-trained controllers}  \label{sec:tra2Dte3D}
In general, a controller trained for a 2D flow should not be expected to perform well for a 3D flow. However, given the efficacy of  DPM-trained controllers for 2D flows (Section~\ref{sec:2D}), it is natural to investigate their potential for 3D flow control, particularly given their relatively low training cost and the largely two-dimensional nature of the flow in the vicinity of the controller, particularly for pre-stall angles of attack. We therefore investigate the effectiveness of 2D-trained controllers for 3D flow control, where both the flow regime (3D) and test time ($t_\mathrm{test}=30$) are out-of-sample.

\begin{figure}
\centering
\includegraphics[width=1\linewidth,trim={0 5 0 0},clip]{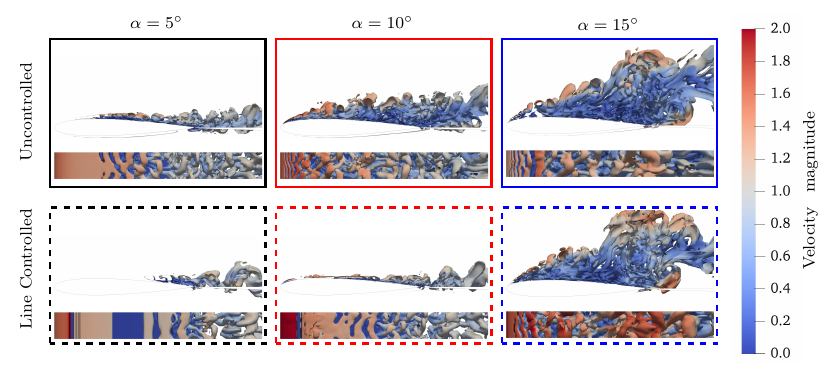}
\caption{{Instantaneous Q-criterion of 3D uncontrolled and controlled airfoil flows. The controlled flows use 2D-trained controllers.}}\label{fig:2Dtra3Dte_Qcriterion}
\end{figure}

\figref{fig:2Dtra3Dte_Qcriterion} shows the instantaneous Q-criterion for 3D uncontrolled and controlled flows for $\alpha=5^\circ$, $10^\circ$, and $15^\circ$, where the controlled flows use LC2 controllers.
At $\alpha=5^\circ$, the controller creates a pronounced, line-shaped suction slot that significantly delays the transition to turbulence, with similar overall effectiveness to the controlled 2D flows (Section~\ref{sec:trate2D_flow} and~\cite{Huang2004numerical}).
At $\alpha=10^\circ$, the controller substantially depresses separation, with the controlled flow accelerating at the leading edge and remaining attached  downstream. 
At $15^\circ$, the controller is less effective, with its influence limited to a slight reduction in leading-edge vortex breakdown.
 
\begin{figure}
 \centering
\includegraphics[width=1.0\linewidth, trim=5 5 0 0, clip]{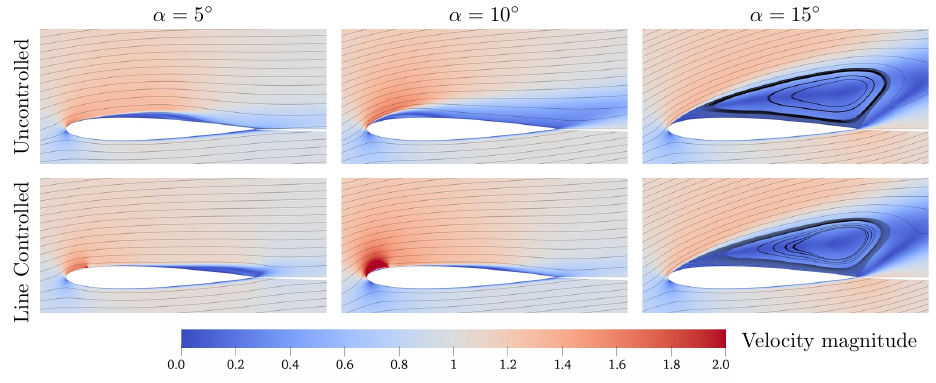}
\caption{{Reynolds-averaged velocity magnitudes and velocity streamlines of 3D uncontrolled and controlled for airfoil flows. The controlled flows use 2D-trained controllers.}}\label{fig:2Dtra3Dte_Streamline}
\end{figure}

\begin{figure}
 \centering
\includegraphics[width=1.0\linewidth, trim=0 10 0 0, clip]{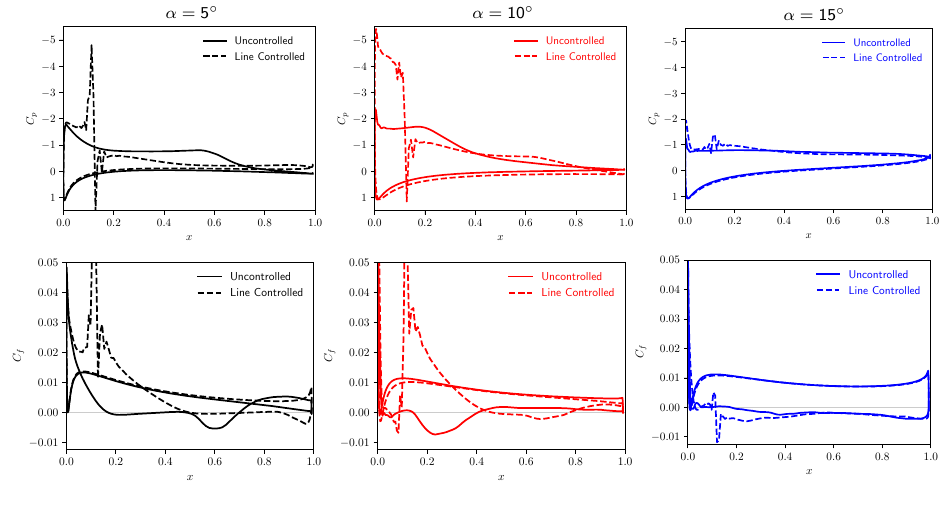}
\caption{{Reynolds-averaged pressure and friction coefficients of 3D airfoils flows; controlled flows use 2D-trained controllers.}}\label{fig:2Dtra3Dte_Cp_Cf_cmp}
\end{figure}

Flow separation can be inferred from the Reynolds-averaged velocity magnitude and streamlines shown in \figref{fig:2Dtra3Dte_Streamline} and the $C_f$  distributions shown in \figref{fig:2Dtra3Dte_Cp_Cf_cmp}. A sudden drop of $C_f$ or the beginning of a flat region identifies the separation point. For example, the separation point of uncontrolled $\alpha=5^\circ$ flow is at $x_\mathrm{sep}=0.16$, with nearly constant $C_f=0$ for $0.16<x<0.71$ indicating  flow separation. In the uncontrolled flows in \figref{fig:2Dtra3Dte_Cp_Cf_cmp}, an elbow-shaped negative region of $C_f$ indicates the transition to turbulence.

For the $\alpha=5^\circ$ flow, the line controller delays the separation visible in the uncontrolled flow (\figref{fig:2Dtra3Dte_Streamline}), particularly near the trailing edge. Since the 3D $\alpha=5^\circ$ leading-edge flow closely resembles that of the 2D flow, with no spanwise vortex present, the line controller's inputs are similar to its 2D counterpart flow, which results in effective control.

For the $\alpha=10^\circ$ flow, the controller significantly depresses separation, yielding a much thinner separation bubble.
However, as for the $\alpha=15^\circ$ cases, the LC2 controllers have a negligible impact on the 3D stalled flow except for minor perturbations on $C_f$ and $C_p$ near the jet. This is understandable, as the 3D post-stall flow is fully separated and turbulent, which leads to rapid dissipation of control energy. Consequently, the control efforts learned for 2D flows are ineffective for the stalled flow, which highlights the challenge applying 2D-trained models to control fully separated turbulent flows.

As shown in the $C_p$ plots (Figure~\ref{fig:2Dtra3Dte_Cp_Cf_cmp}), all controllers generate a pressure spike near the jet and partially or fully mitigate the adverse pressure region on the upper surface of the airfoil. The influence of line controllers is primarily global and pronounced, resulting in significant changes to both the minimum $C_p$ and the pressure differential between the upper and lower surfaces. These changes substantially affect drag and lift performance (discussed subsequently in Section~\ref{sec:trate3D_CdCl}). Notably, the $\alpha = 5^\circ$ flow remains largely two-dimensional at the leading edge, leading to $C_p$ distributions from the line controllers that closely resemble those in 2D flows. Similarly, the 3D $\alpha = 10^\circ$ flow exhibits $C_p$ patterns comparable to the 2D flow, except farther downstream.

\subsection{Control performance of 3D-trained controllers} \label{sec:trate3D}

We now train DPM flow controllers to minimize the objective function (\ref{eq:loss}) over two chord-flow-time intervals ($t_\mathrm{train} = \makeFG{2}$) with simulation time step $\Delta t = 4 \times 10^{-5}$, DPM optimization interval $\tau = 100\Delta t$, and $N_\mathrm{itr} = 500$ optimization iterations. We then test the trained controllers over 30 chord-flow times ($t_\mathrm{test} = 30$) at three different angles of attack, \makeFG{where the tests are out-of-sample in time}. The optimization window timescale is 1.4 times the sonic timescale, based on the distance from the controller to the sensor, thus control-jet pressure waves are able to to propagate to the sensor within one optimization window.

\begin{figure}
 \centering
\includegraphics[width=1.0\linewidth]{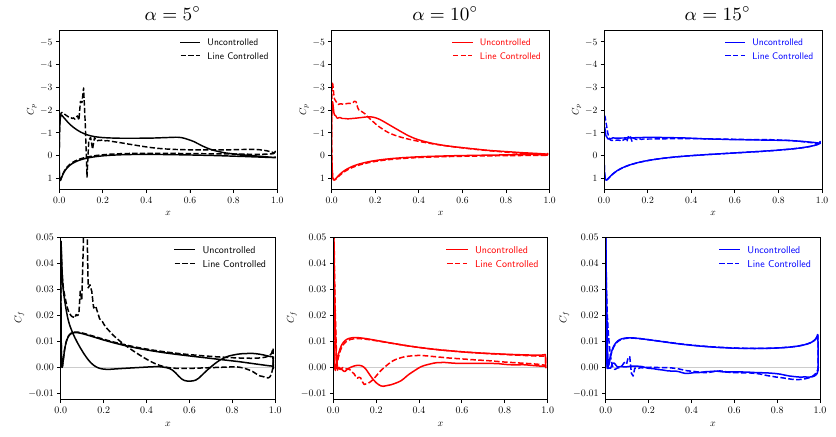}
\caption{Reynolds-averaged pressure and friction coefficients of uncontrolled and controlled 3D airfoil flows using 3D-trained controllers.}\label{fig:3Dtrate_Cp_Cf_cmp}
\end{figure}

For all angles of attack, line controllers significantly alter $C_p$, as shown in \figref{fig:3Dtrate_Cp_Cf_cmp}, especially for pre-stall flows. Notably, the peak negative pressure coefficient is reduced from -1.8 to -3.0 at $\alpha=5^\circ$ and from -2.5 to -3.2 at $\alpha=10^\circ$. For flows at $\alpha=5^\circ$, which are primarily two-dimensional in the vicinity of the controller, the negative pressure zone is globally affected in similarity to the controlled 2D flows (Section~\ref{sec:trate2D_flow}). 
In contrast, for flows at $\alpha=10^\circ$, the negative pressure zone around the leading edge expands, eventually recovering to the uncontrolled flow state downstream of the jet.
The post-stall case at $\alpha=15^\circ$ presents more complexity due to strong nonlinear effects and large-scale flow separation, both of which limit the effectiveness of control in reducing drag.

Flow separation is illustrated by the Reynolds-averaged $C_f$ distributions in \figref{fig:3Dtrate_Cp_Cf_cmp}. For flows at $\alpha=5^\circ$, the controller effectively delays separation, with the separation point shifting downstream to approximately $x = 0.5$, and promotes attached flow toward the trailing edge. At $\alpha = 10^\circ$, the controller suppresses the separation bubble and improves flow attachment compared to the baseline case. At $\alpha=15^\circ$, the controller exhibits minimal influence on the flow separations due to the limited control authority over the highly separated flow.

\begin{figure}
  \centering
\includegraphics[width=0.8\linewidth,trim={0 5 0 0},clip]{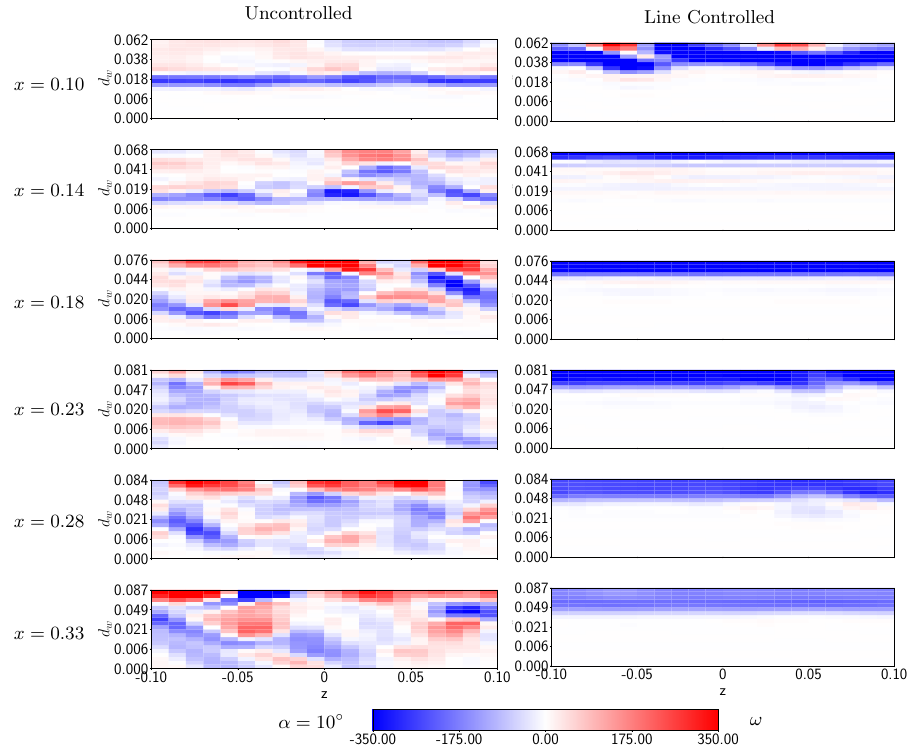}
  \caption{{Instantaneous vorticity magnitude shown in cross-stream ($y$--$z$) planes at six downstream positions ($x$) for  $\alpha = 10^\circ$ uncontrolled and controlled cases. $d_w$ is the distance from the airfoil's upper surface.}}\label{fig:omega}
\end{figure}

To further investigate the effect of the learned controller, \figref{fig:omega} displays instantaneous vorticity magnitude ($\omega = |\nabla\times\bu|$) in cross-stream ($y$--$z$) slices at six streamwise positions $x\in\{0.10,0.14,0.18,0.23,0.28,0.33\}$ for the challenging, partially separated $\alpha=10^\circ$ uncontrolled and controlled flows.
In the uncontrolled case, flow separation occurs before $x=0.18$ and is followed by downstream reattachment. At this angle of attack, strong mixing and enhanced vorticity magnitude are observed between the freestream layer and the boundary layers after the separation point.

In contrast, the line controller generates a significant amount of vorticity and substantially alters the freestream flow near the jet. This vortex is attributed to the adaptive suction that maintains attached flow along much of the airfoil's upper surface. The boundary layer thickness increases compared to the baseline, which demonstrates the actuator's effectiveness. Downstream, the vorticity evolves and diffuses gradually.

\subsection{Lift and drag for 2D- and 3D-trained controllers} \label{sec:trate3D_CdCl}

We now analyze the lift and drag characteristics of the adaptive and constant-pressure controllers across the range of angles of attack assessed previously, from attached flow ($\alpha=5^\circ$) to partially separated ($\alpha=10^\circ$) and fully separated ($\alpha=15^\circ$) flow. As we show, the adaptive controller trained for 3D airfoil flows (LC3) is the only controller that consistently improves the $\clcd$ and $\clcdc$ targets, while its constant-pressure counterpart (LC3c) and the 2D-trained controllers exhibit inferior performance and/or undesirable controlled-flow characteristics.

\begin{figure}
 \centering
\includegraphics[width=0.33\linewidth, trim={10 0 10 0}, clip]{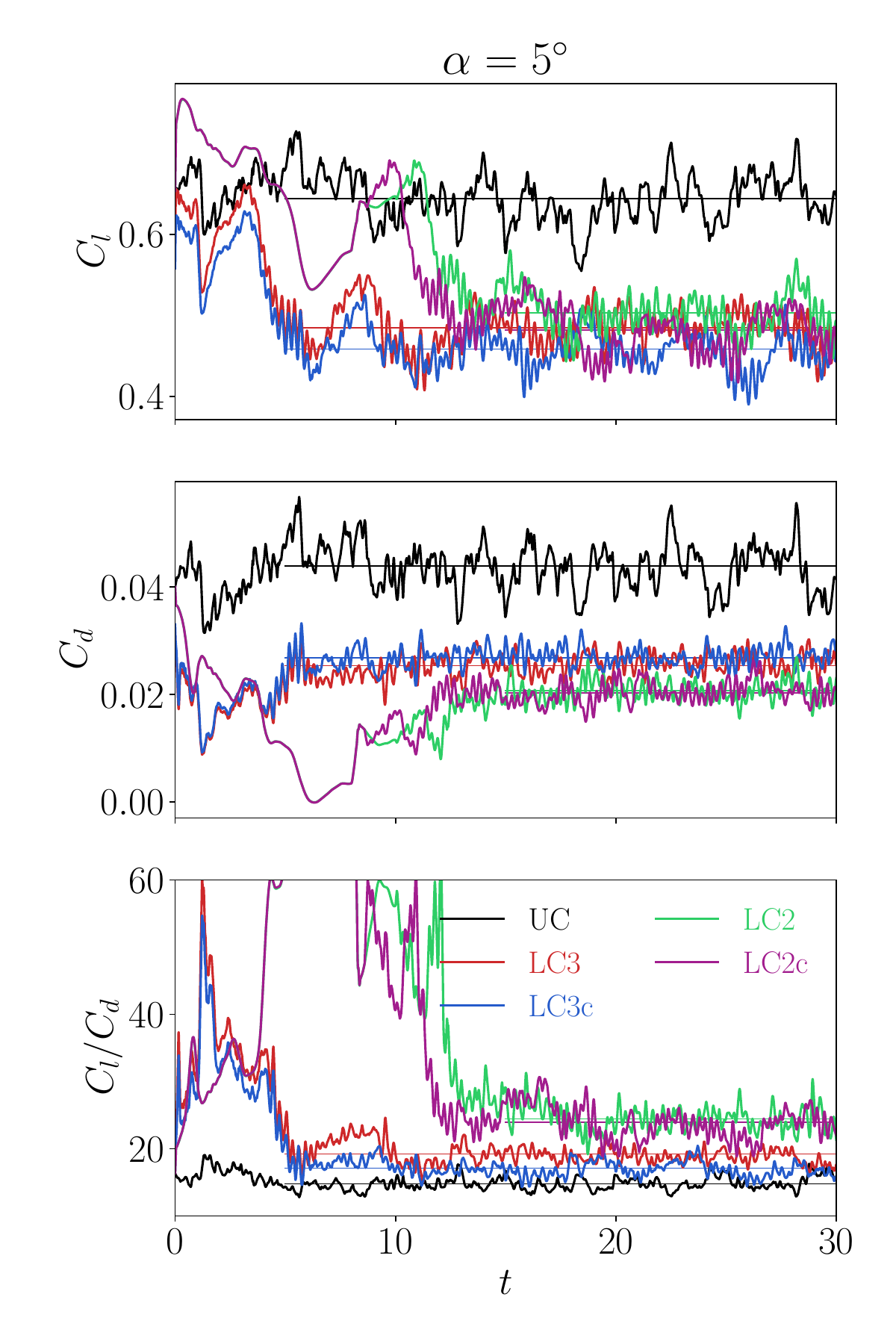}
\includegraphics[width=0.33\linewidth, trim={10 0 10 0}, clip]{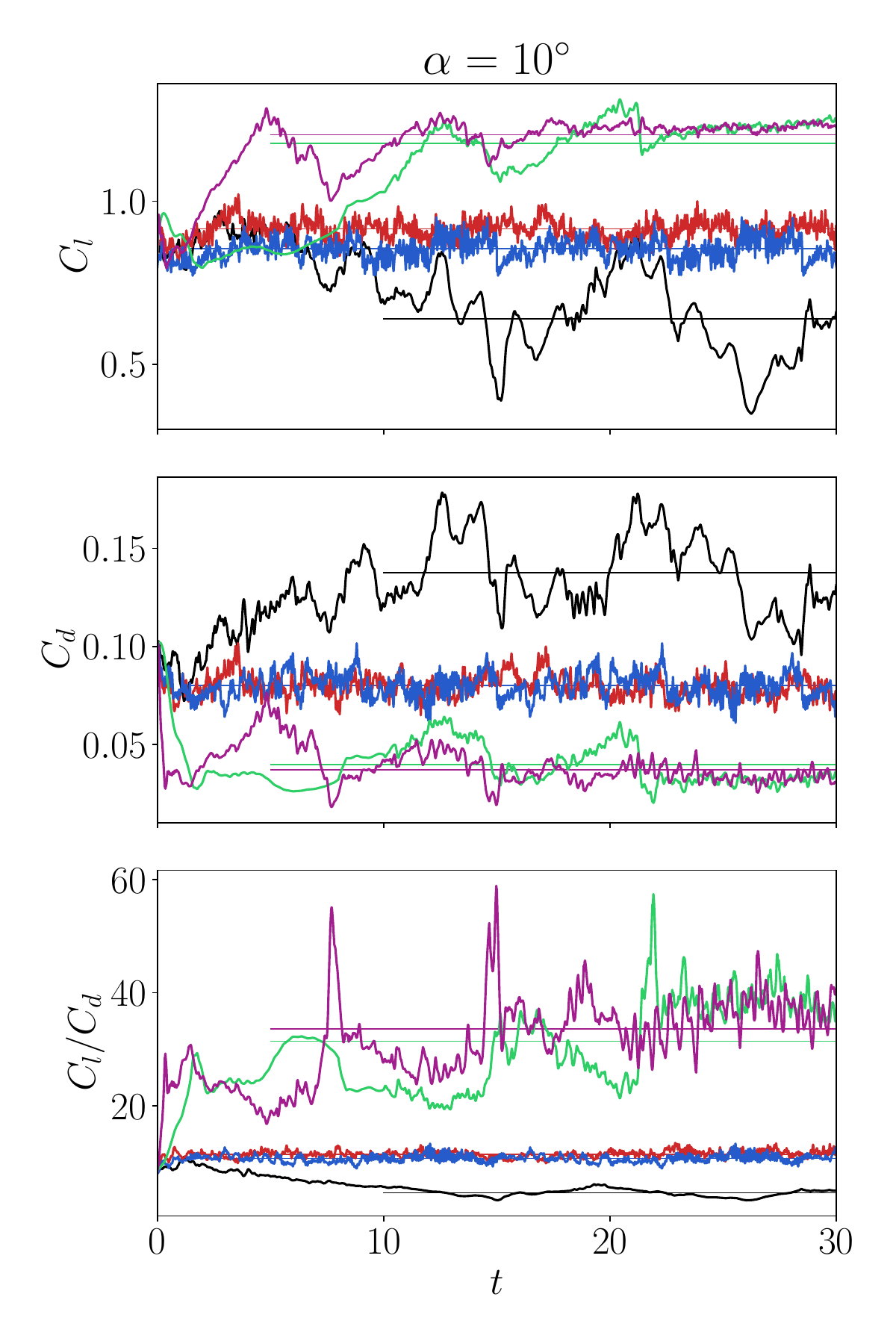}
\includegraphics[width=0.33\linewidth, trim={10 0 10 0}, clip]{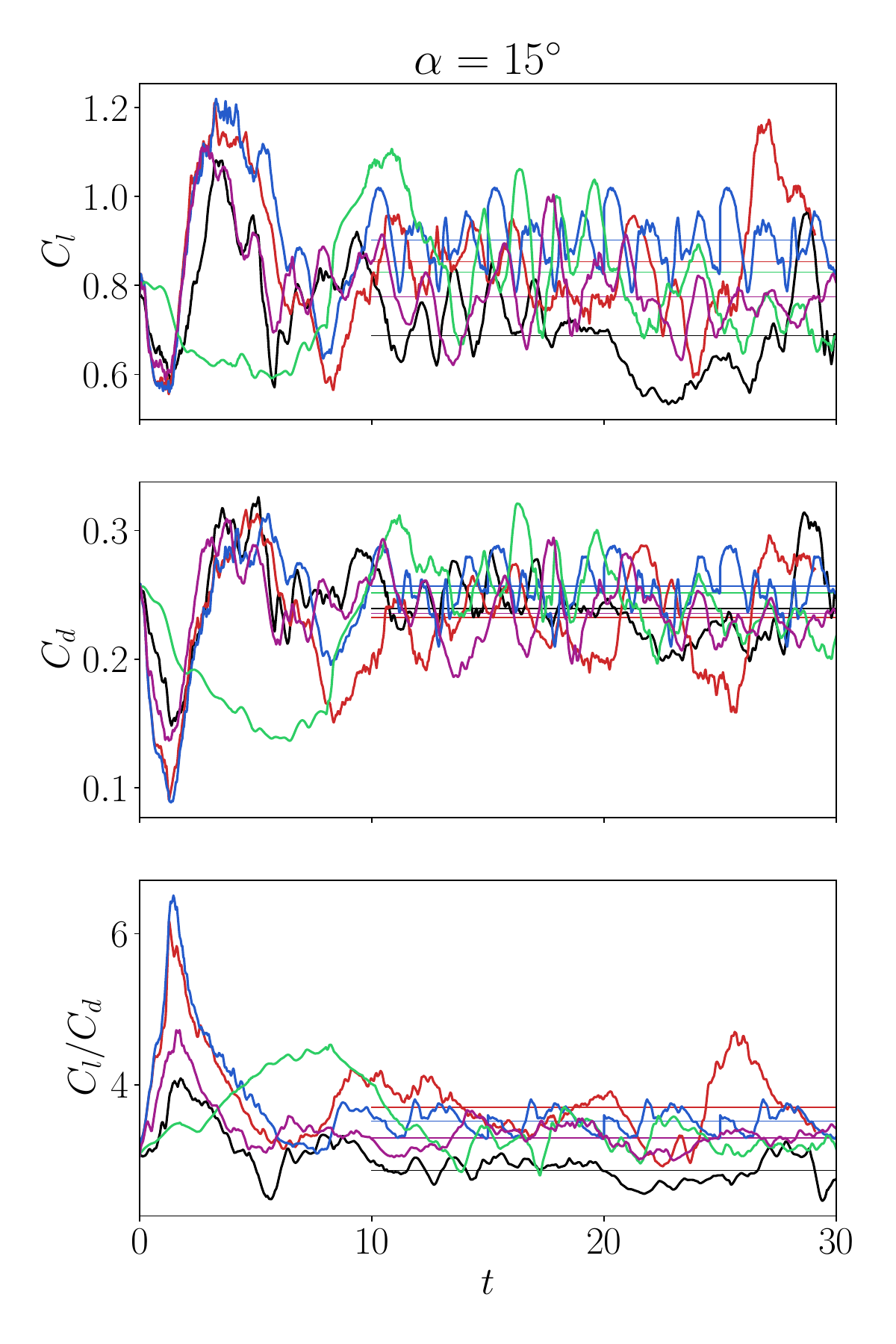}
\caption{{Instantaneous and time-averaged lift coefficient, drag coefficient, and lift-to-drag ratio for 3D uncontrolled (UC) and controlled (LC) turbulent airfoil flows.}}\label{fig:3D_tClCdr}
\end{figure}

\begin{figure}
\centering
\includegraphics[width=0.8\linewidth, trim={0 0 0 60}, clip]{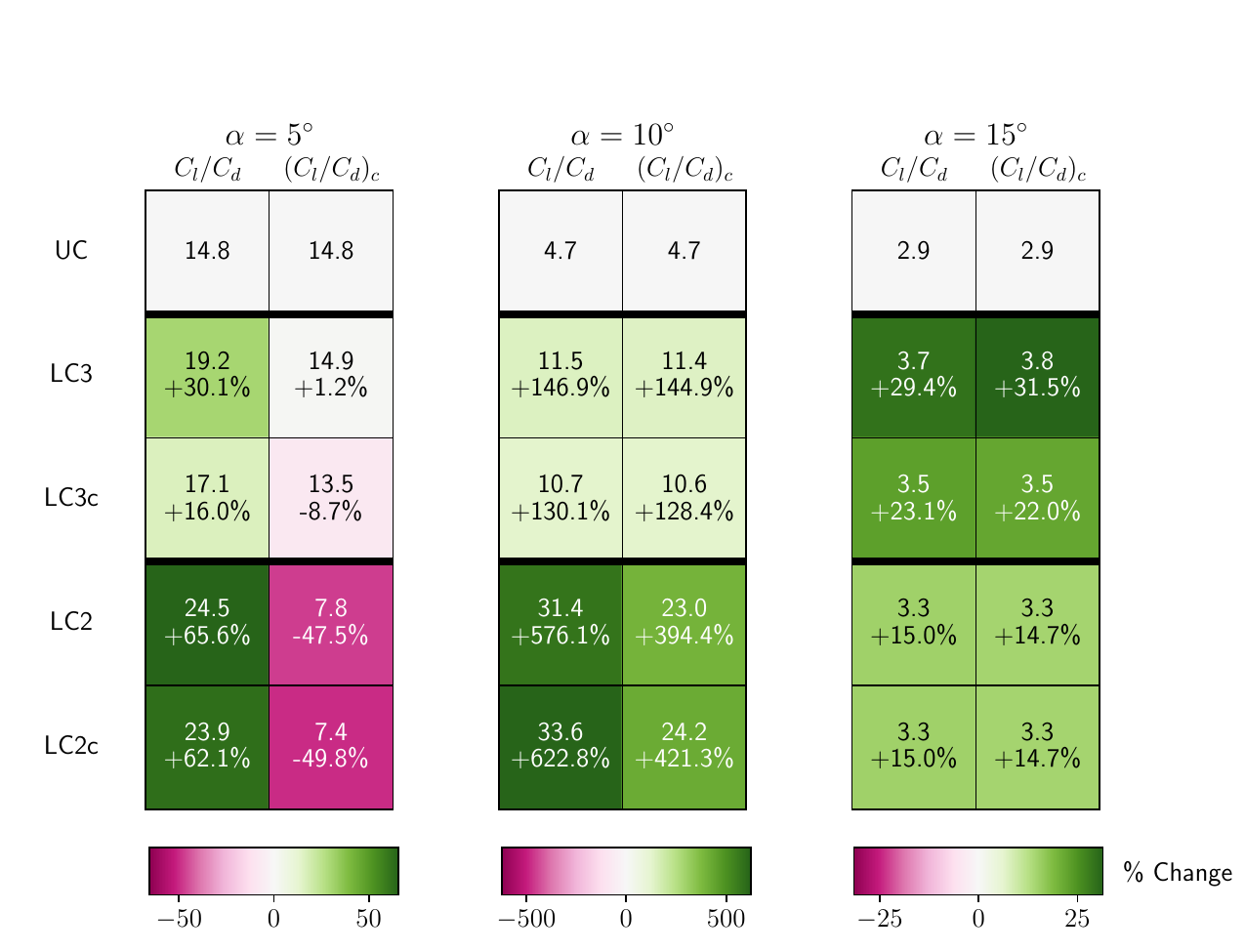}
\caption{{Lift-to-drag and energy-corrected lift-to-drag ratios of uncontrolled and controlled 3D airfoil flows using 2D- and  3D-trained adaptive controllers (LC2 and LC3) and their time-averaged, constant-control counterparts (LC2c and LC3c). Percent changes are relative to the uncontrolled (UC) flow.
}}\label{fig:2Dvs3D}
\end{figure}

\figref{fig:3D_tClCdr} presents the instantaneous $C_l$, $C_d$, and $C_l/C_d$ for uncontrolled (UC) flows and controlled flows using 2D- and 3D-trained adaptive controllers (LC2 and LC3) and their time-averaged, constant-control counterparts (LC2c and LC3c). As could be expected by their marked effect on the pressure and friction coefficients shown previously, all controllers quickly modify both lift and drag at the onset of control. Figure~\ref{fig:2Dvs3D} summarizes the time-averaged lift-to-drag ratio ($C_l/C_d$) and the energy-corrected lift-to-drag ratio $(C_l/C_d)_c$, as well as the percent change in each (relative to the uncontrolled flow) for the present 3D turbulent flows.

At $\alpha = 5^\circ$, LC3 achieves a 30\,\% improvement in $C_l/C_d$ over the uncontrolled flow, primarily through drag reduction, and stabilizes the flow shortly after $t_\mathrm{test} = 5$ as it adapts to spanwise flow dynamics. While the resulting 1.2\,\% improvement in $\clcdc$ is small, it is important to note that LC3 is the \emph{only} controller able to improve $\clcdc$ for the $\alpha=5^\circ$ flow. Its time-averaged counterpart, LC3c,  improves $\clcd$ by a modest 16\,\% but does so with reduced overall system efficiency, exhibiting an 8.7\,\% reduction in $\clcdc$.

Since the $\alpha=5^\circ$ flow is entirely laminar in the vicinity of the slot-jet controller, the 2D-trained LC2 and LC2c controllers are also effective at improving $\clcd$, giving 65.6\,\% and 62.1\,\% improvement, respectively. However, their high actuation pressures (discussed subsequently in Section~\ref{sec:2Dtravs3Dtra}) cause them to operate at a severe energy penalty, with system-level $\clcdc$ reductions of nearly 50\,\%. Their overly-strong actuation is apparent in the large $\clcd$ transients they cause when initially applied to the uncontrolled flow (see Figure~\ref{fig:3D_tClCdr}).

At $\alpha = 10^\circ$, the flow structure's marginal stability  causes the uncontrolled flow to oscillate between partial and full separation---we note that even DNS calculations disagree on whether the time-averaged flow structure is fully separated~\cite{Rodriguez2013,hickling2024large} or reattaching~\cite{Turner2020a} around this angle of attack. 
The LC3 controller, trained for these 3D flow dynamics, is able to fully stabilize this process and achieve 147\,\% improvement in $\clcd$ and 145\,\% improvement in $\clcdc$ over the baseline, while its time-averaged counterpart (LC3c) is 17\,\% less effective at improving $\clcd$ and $\clcdc$ due to its lack of fine-scale adjustments, which we discuss subsequently in Section~\ref{sec:2Dtravs3Dtra}. This demonstrates one advantage of the adaptive LC3 controller over the constant-pressure LC3c.

At $\alpha=10^\circ$, the 2D-trained LC2 controller (and its constant-control counterpart LC2c) actuate the flow dramatically, increasing $\clcd$ by a factor of six over the baseline flow, with a concomitant increase of $\clcdc$ by a factor of four. The spread between $\clcd$ and $\clcdc$ for the 2D-based controllers is much greater than that of the 3D-based controllers for this angle of attack, which indicates dimishing returns (i.e., increasing energy penalty) due to overactuation. Additionally, unlike the LC3-controlled flow, the LC2-controlled flow exhibits strong lift and drag oscillations (see Figure~\ref{fig:3D_tClCdr}) that would likely be disadvantageous in flight. LC2 was not trained for the 3D boundary-layer separation and reattachment dynamics that the $\alpha=10^\circ$ case exhibits and therefore cannot suppress these large transients as LC3 can.

Despite the flow being fully separated at $\alpha=15^\circ$, LC3 achieves a significant 29\,\% improvement of $C_l/C_d$ and a corresponding  31\,\% improvement of $\clcdc$. As with the $\alpha=5^\circ$ and $10^\circ$ flows, LC3c does not fare as well as LC3, here offering 23\,\% improvement of $\clcd$ and 22\,\% improvement of $\clcdc$. The 2D-trained LC2 controller and its constant LC2c counterpart provide approximately half the improvement of LC3 for this case, likely due to the fact that LC3 optimizes over this case's fully separated, turbulent dynamics during its training, while LC2 does not.

Across the range of angles of attack tested, LC3 is the only controller that consistently improves both $\clcd$ and the energy-penalized $\clcdc$. This is consistent with the training of LC3 to minimize the energy-penalized objective function \eqref{eq:loss} over the present 3D turbulent flow dynamics. LC3's constant-pressure counterpart (LC3c) does not adaptively force the flow and so underperforms LC3 in all cases.
The 2D-trained controllers (LC2 and LC2c) are not optimized over the fully 3D turbulent flow and so either overactuate the flow ($\alpha=5^\circ$), incurring excessive energy penalties, lead to the formation of large-amplitude, transient disturbances ($\alpha=10^\circ$), or underactuate the fully separated turbulent flow ($\alpha=15^\circ$). These overall trends underscore the advantages of the adaptive, 3D-trained LC3 controller.
We next analyze the control maps of the learned adaptive controllers to further understand the reasons for LC3's overall leading performance.

\subsection{Further analysis of adaptive and constant-pressure actuators}
\label{sec:2Dtravs3Dtra}

\begin{figure}
\centering
\includegraphics[width=0.33\linewidth, trim={10 0 10 0}, clip]{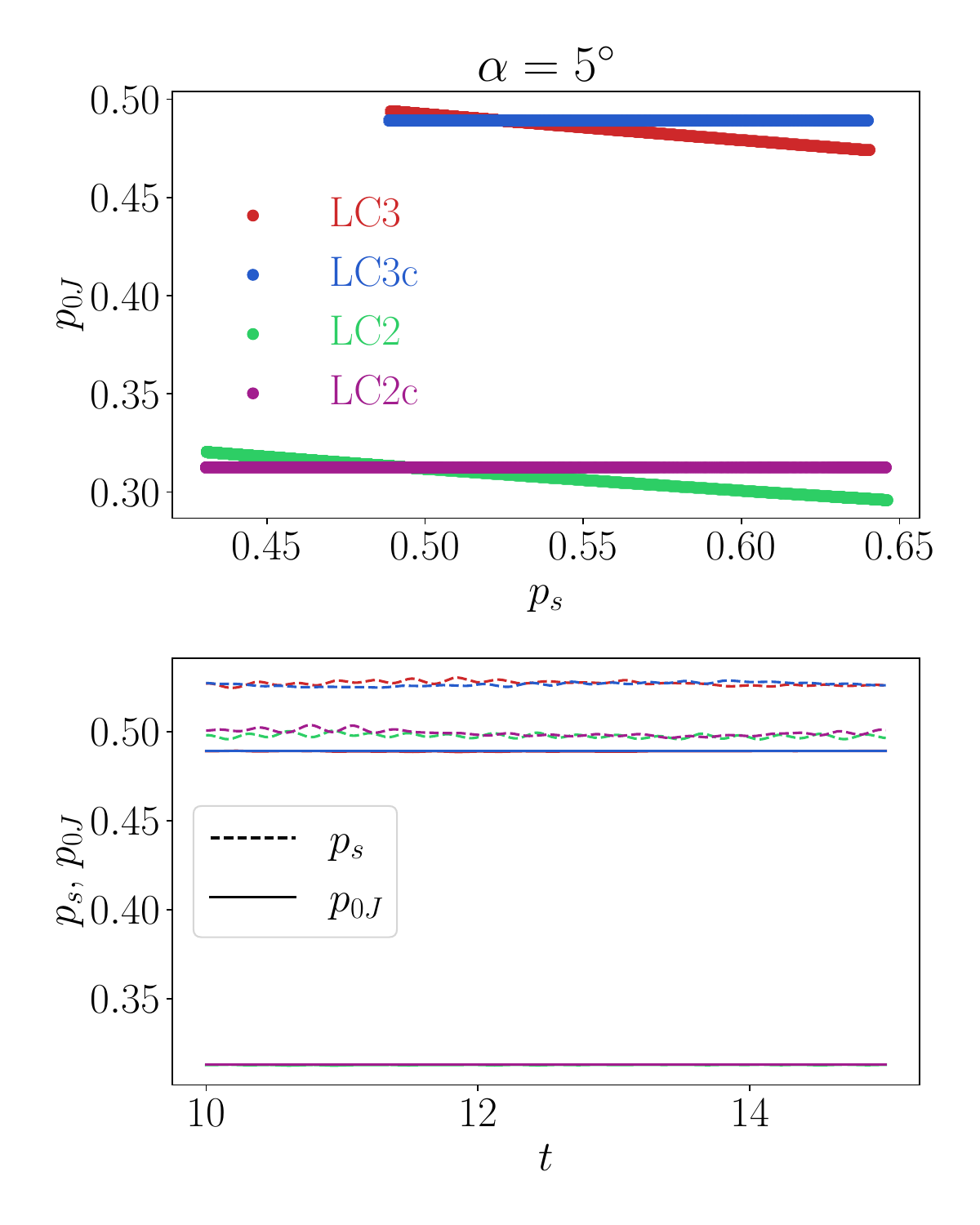}
\includegraphics[width=0.33\linewidth, trim={10 0 10 0}, clip]{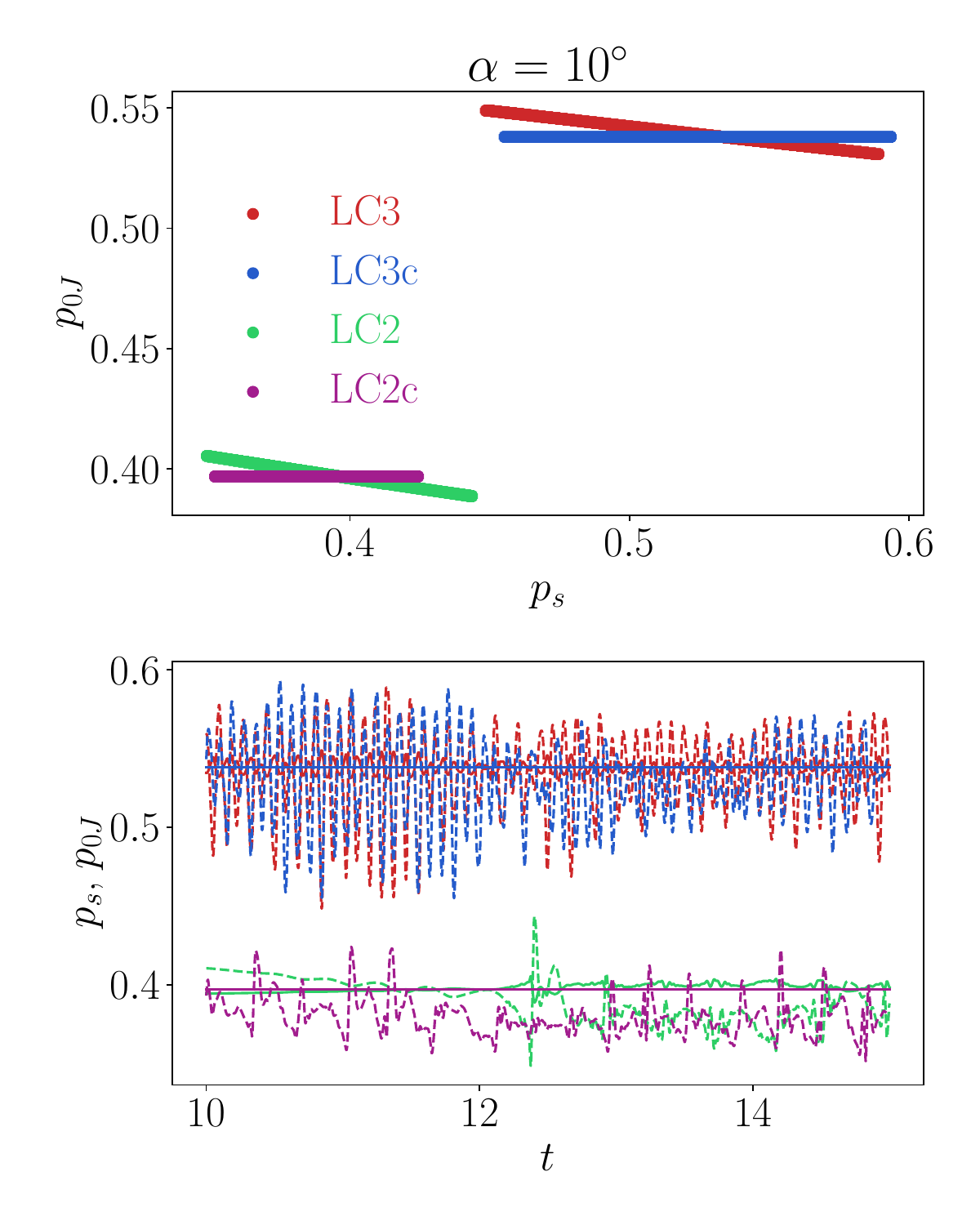}
\includegraphics[width=0.33\linewidth, trim={10 0 10 0}, clip]{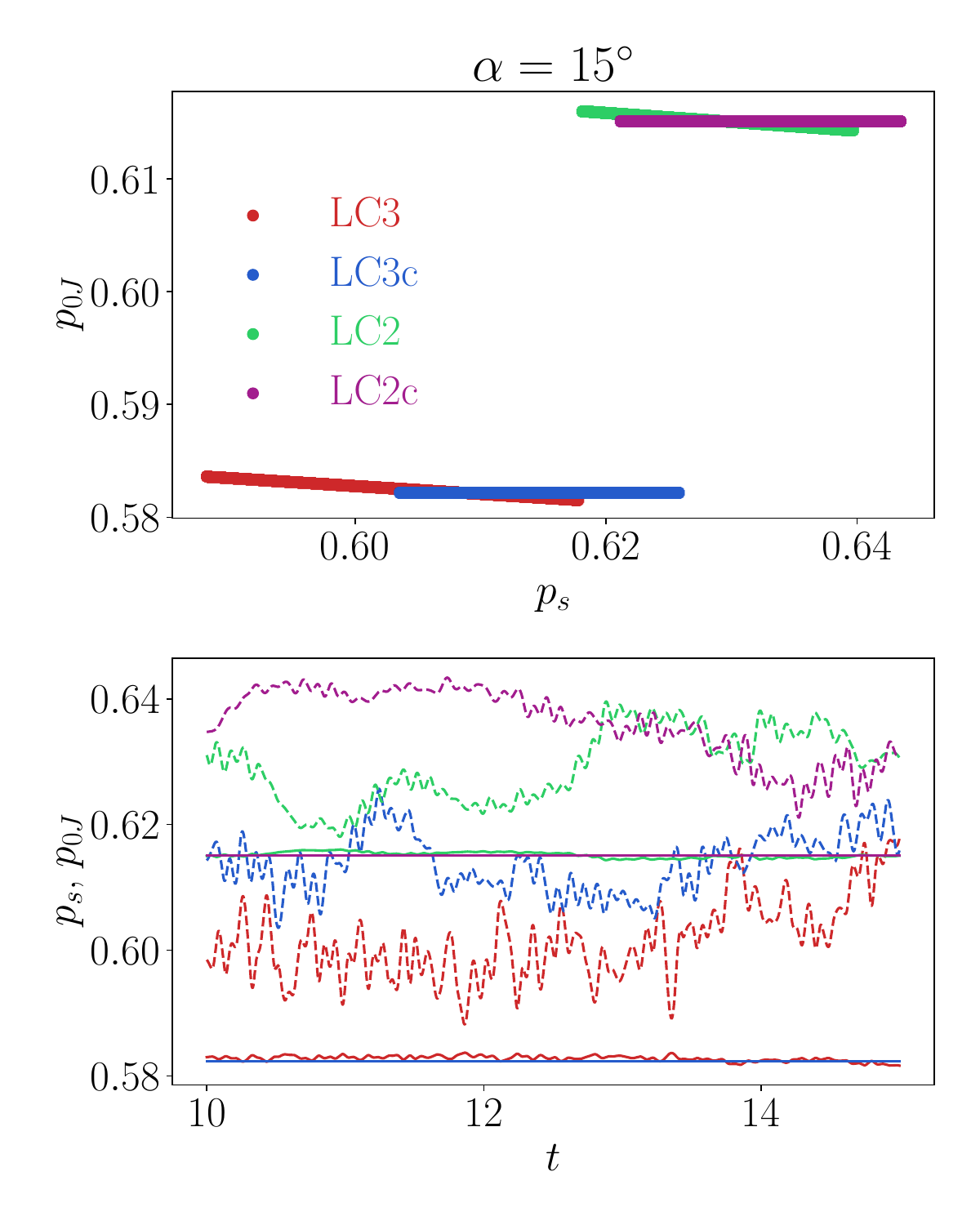}
\caption{{Top row: Control maps of 2D- and 3D-trained adaptive and constant controllers evaluated for 3D airfoil flows over $t_\mathrm{test}\in[0,30]$. Bottom row: instantaneous sensed pressure $p_s$ and commanded jet total pressure $p_{0J}$ for a sample window $t_\mathrm{test}\in[10,15]$.}}\label{fig:psvsp0j}
\end{figure}

To further analyze the reasons for the various controllers' relative performance, Figure~\ref{fig:psvsp0j} plots control maps of the commanded jet total pressure ($p_{0J}$) versus the sensed airfoil surface pressure ($p_s$), evaluated \emph{in situ} during \emph{a posteriori} simulations of controlled turbulent airfoil flows, and example $p_s$ and $p_{0J}$ time traces for $t_\mathrm{test}\in[10,15]$. In these plots, lower $p_{0J}$ corresponds to stronger actuation (stronger suction) and higher control power $P_J$ \eqref{eq:control_power}.

At $\alpha = 5^\circ$ and $10^\circ$,
LC2 commands stronger suction than LC3, which leads to the former's overall stronger actuation and larger $C_l/C_d$ changes---and concomitantly larger energy penalties (larger gap between $\clcd$ and $\clcdc$)---as shown in Figure~\ref{fig:2Dvs3D}. In contrast, for the post-stall $\alpha=15^\circ$ case, LC3 learns stronger suction than LC2, which is necessary for this fully separated flow and leads to LC3's overall more consistent control performance. 

The attached, laminar boundary layer of the $\alpha=5^\circ$ flow is apparent in Figure~\ref{fig:psvsp0j}, in which both $p_s$ and $p_{0J}$ are virtually constant in time.
At $\alpha=10^\circ$, rapid fluctuations in the sensed pressure cause LC3 to command micro-adjustments about the mean $p_{0J}$, which results in its greater improvement of $\clcd$ and $\clcdc$ than LC3c, while LC2 (not trained for the 3D flow dynamics) does not exhibit the same fine-scale adjustments as its larger applied suction suppresses $p_s$ fluctuations. 
LC3 also exhibits these micro-adjustments at $\alpha=15^\circ$, for which it more effectively reduces the upper-surface pressure than the other controllers and hence best improves $\clcd$. Importantly, even at this post-stall angle of attack,  LC3's dynamical actuation of the turbulent boundary layer improves the target $\clcd$ and $\clcdc$ metrics more effectively than its constant-pressure counterpart (LC3c), which further highlights the advantage of adaptive control over constant-pressure control.

\section{Conclusion} \label{sec:conclusion}

We develop deep learning active flow controllers for NACA 0012 airfoil flows using a blowing/suction slot jet along the airfoil's upper surface. The control actuation is optimized to increase the lift-to-drag ratio with a penalty on excessive control power. The optimization is performed simultaneously with the solution of the controlled PDEs using adjoints of the flow variables to provide the necessary end-to-end sensitivities. This \emph{in situ} optimization enables accurate, dynamical response of the deep learning controllers to pressure transients, which leads to greater control effectiveness than the corresponding constant-total pressure jets. Previous work has compared the performance of adjoint-trained deep learning flow controllers to deep reinforcement learning-based controllers for laminar flows; this work extends the adjoint-based approach to control of unsteady and turbulent airfoil flows.

For 2D flows, the adjoint-based controllers successfully modify the vortex shedding to delay flow separation at pre-stall angles of attack, leading to 95\,\% increase of the lift-to-drag ratio at $\alpha=5^\circ$ and 171\,\% increase of the lift-to-drag ratio at $\alpha=10^\circ$. While the control efficiency is relatively low at $\alpha=5^\circ$ (corrected aerodynamic efficiency improvement of 3\,\%), the corrected aerodynamic efficiency increase of 132\,\% at $\alpha=10^\circ$ is consistent with its actual aerodynamic performance increase, indicating efficient control. Control efficiency at post-stall $\alpha=15^\circ$ is notable (34\,\% improvement of corrected aerodynamic efficiency). We also analyze the effect of optimization window size using the 2D flows, finding the optimum to be $\tau\in[5\Delta t, 200\Delta t]$ simulation time steps  ($\tau \in [1\times10^{-3}, 4\times 10^{-4}]$ chord flow times) for these flows.

In general, a controller trained for a 2D flow can not be expected to perform well for a 3D flow. When applying 2D-trained active controllers to out-of-sample 3D flows, the spanwise locally actuated line controller creates a pronounced suction slot that significantly delays the transition to turbulence, increasing $\clcd$ for all angles of attack, though for pre-stall 3D flows, these controllers' overactuation leads to significant actuation-power penalties, and for the post-stall $\alpha=15^\circ$ flow, their underactuation leads to lower effectiveness than the 3D-trained controllers.

When trained directly for 3D turbulent flows, the deep learning adaptive flow controllers further enhance aerodynamic performance by suppressing flow separation, increasing $C_l/C_d$, and improving the power-weighted $(\clcd)_c$.
Significantly, the 3D-trained adaptive controller (LC3) is the only controller that consistently improves these performance metrics across all angles of attack tested, with constant-control-pressure analogs (LC3c) having reduced control effectiveness and efficiency at all angles of attack.
These trends underscore the advantages of the adaptive, DPM-based control strategy in identifying and implementing effective and efficient control of turbulent airfoil flows.

The success of the learned active controllers highlights the importance of adaptive control policies compared to fixed control outputs. By optimally mapping local pressure measurements to real-time actuation, the DPM-based control framework enables sensor-based policies to respond to changing flow conditions across different flow regimes. While constant-pressure control may be effective in narrowly defined cases, the DPM-based control strategy offer a robust, efficient, and transferable solution for high-dimensional, dynamically evolving flow environments.

The effectiveness of adjoint-based deep learning for unsteady and turbulent airfoil flows is promising and offers valuable insights into control methodologies for increasingly complex flows. It shows significant potential for advancing dynamic separation and stall control, especially for gusting conditions and pitching airfoils.
In compressible regimes, the adjoint-based control approach is expected to address challenges related to shock-wave interactions and high-speed flow instabilities, potentially enabling more precise control of shock-boundary layer interactions at high Mach numbers.
For highly nonlinear flows, such as reacting flows, adjoint-based deep learning could provide an efficient means to accurately model and control complex chemical reactions and turbulence--chemistry interactions, paving the way for improved control and optimization in demanding combustion systems.

\section*{Acknowledgments}

The authors are grateful to Justin Sirignano for many helpful comments and suggestions. This material is based in part upon work supported by the U.S.\ National Science Foundation under Award CBET-22-15472. 
This work is supported by the U.K.\ Engineering and Physical Sciences Research Council grant EP/X031640/1.
This research used resources supported by the University of Notre Dame Center for Research Computing and resources of the Oak Ridge Leadership Computing Facility, which is a DOE Office of Science User Facility supported under Contract DE-AC05-00OR22725.


\begin{thebibliography}{70}
\newcommand{\enquote}[1]{``#1''}
\providecommand{\natexlab}[1]{#1}
\providecommand{\url}[1]{\texttt{#1}}
\providecommand{\urlprefix}{URL }
\expandafter\ifx\csname urlstyle\endcsname\relax
  \providecommand{\doi}[1]{\discretionary{}{}{}https://doi.org/#1}\else
  \providecommand{\doi}[1]{\discretionary{}{}{}\urlstyle{rm}\url{https://doi.org/#1}}\fi

\bibitem[{Dussauge et~al.(2023)Dussauge, Sung, Pinon~Fischer, and
  Mavris}]{Dussauge2023reinforcement}
Dussauge, T.~P., Sung, W.~J., Pinon~Fischer, O.~J., and Mavris, D.~N.,
  \enquote{A Reinforcement Learning Approach to Airfoil Shape Optimization,}
  \emph{Scientific Reports}, Vol.~13, No.~1, 2023, p. 9753.
\newblock \doi{10.1038/s41598-023-36560-z}.

\bibitem[{Skinner and Zare-Behtash(2018)}]{Skinner2018state}
Skinner, S., and Zare-Behtash, H., \enquote{State-of-the-Art in Aerodynamic
  Shape Optimisation Methods,} \emph{Applied Soft Computing}, Vol.~62, 2018,
  pp. 933--962.
\newblock \doi{10.1016/j.asoc.2017.09.030}.

\bibitem[{Rabault et~al.(2019)Rabault, Kuchta, Jensen, Réglade, and
  Cerardi}]{Rabault2019artificial}
Rabault, J., Kuchta, M., Jensen, A., Réglade, U., and Cerardi, N.,
  \enquote{Artificial Neural Networks Trained through Deep Reinforcement
  Learning Discover Control Strategies for Active Flow Control,} \emph{Journal
  of Fluid Mechanics}, Vol. 865, 2019, p. 281–302.
\newblock \doi{10.1017/jfm.2019.62}.

\bibitem[{{\"O}zkan(2022)}]{ozkan2022active}
{\"O}zkan, M., \enquote{Active and Passive Flow Control Methods over Airfoils
  for Improvement in Aerodynamic Performance,} \emph{New Frontiers in
  Sustainable Aviation}, Springer, 2022, pp. 19--33.
\newblock \doi{10.1007/978-3-030-80779-5_2}.

\bibitem[{Satar et~al.(2024)Satar, Razak, Abdullah, and Ismal}]{satar2024a}
Satar, M. H.~A., Razak, N., Abdullah, M.~S., and Ismal, F., \enquote{A
  Comprehensive Comparison of Passive Flow Controls on the Wind Turbine Blade
  Lift and Drag Performances: A CFD Approach,} \emph{European Journal of
  Mechanics - B/Fluids}, Vol. 108, 2024, pp. 119--133.
\newblock \doi{10.1016/j.euromechflu.2024.07.009}.

\bibitem[{Gerontakos and Lee(2006)}]{Gerontakos2006dynamic}
Gerontakos, P., and Lee, T., \enquote{Dynamic Stall Flow Control via a
  Trailing-Edge Flap,} \emph{AIAA Journal}, Vol.~44, No.~3, 2006, pp. 469--480.
\newblock \doi{10.2514/1.17263}.

\bibitem[{Bliamis et~al.(2022)Bliamis, Vlahostergios, Misirlis, and
  Yakinthos}]{Bliamis2022numerical}
Bliamis, C., Vlahostergios, Z., Misirlis, D., and Yakinthos, K.,
  \enquote{{Numerical Evaluation of Riblet Drag Reduction on a MALE UAV},}
  \emph{Aerospace}, Vol.~9, No.~4, 2022, p. 218.
\newblock \doi{10.3390/aerospace9040218}.

\bibitem[{Juvinel et~al.(2023)Juvinel, Roa, and
  Schaerer}]{Juvinel2023structural}
Juvinel, J. M. D.~E., Roa, D. P.~P., and Schaerer, C.~E., \enquote{Structural
  and Shape Optimization in Aerodynamic Airfoil Performance: A Literature
  Review,} \emph{Preprints}, 2023.
\newblock \doi{10.20944/preprints202307.0807.v1}.

\bibitem[{He et~al.(2019)He, Li, Mader, Yildirim, and Martins}]{He2019robust}
He, X., Li, J., Mader, C.~A., Yildirim, A., and Martins, J.~R., \enquote{Robust
  Aerodynamic Shape Optimization—From a circle to an Airfoil,}
  \emph{Aerospace Science and Technology}, Vol.~87, 2019, pp. 48--61.
\newblock \doi{10.1016/j.ast.2019.01.051}.

\bibitem[{Jameson(1995)}]{jameson1995optimum}
Jameson, A., \enquote{Optimum Aerodynamic Design Using CFD and Control Theory,}
  \emph{12th AIAA Computational Fluid Dynamics Conference}, 1995, p. 1729.
\newblock \doi{10.2514/6.1995-1729}.

\bibitem[{Nadarajah and Jameson(2001)}]{nadarajah2001studies}
Nadarajah, S., and Jameson, A., \enquote{Studies of the Continuous and Discrete
  Adjoint Approaches to Viscous Automatic Aerodynamic Shape Optimization,}
  \emph{15th AIAA Computational Fluid Dynamics Conference}, 2001, p. 2530.
\newblock \doi{10.2514/6.2001-2530}.

\bibitem[{Srinath and Mittal(2010)}]{Srinath2010an}
Srinath, D., and Mittal, S., \enquote{An Adjoint Method for Shape Optimization
  in Unsteady Viscous Flows,} \emph{Journal of Computational Physics}, Vol.
  229, No.~6, 2010, pp. 1994--2008.
\newblock \doi{10.1016/j.jcp.2009.11.019}.

\bibitem[{Chiba et~al.(2005)Chiba, Obayashi, Nakahashi, and
  Morino}]{Chiba2005high}
Chiba, K., Obayashi, S., Nakahashi, K., and Morino, H., \enquote{High-fidelity
  Multidisciplinary Design Optimization of Aerostructural Wing Shape for
  Regional Jet,} \emph{23rd AIAA Applied Aerodynamics Conference}, 2005, p.
  5080.
\newblock \doi{10.2514/6.2005-5080}.

\bibitem[{{\"U}m{\"u}tl{\"u} and Kiral(2022)}]{umutlu2022airfoil}
{\"U}m{\"u}tl{\"u}, H. C.~A., and Kiral, Z., \enquote{Airfoil Shape
  Optimization Using B{\'e}zier Curve and Genetic Algorithm,} \emph{Aviation},
  Vol.~26, No.~1, 2022, pp. 32--40.
\newblock \doi{10.3846/aviation.2022.16471}.

\bibitem[{Fourie and Groenwold(2002)}]{fourie2002particle}
Fourie, P., and Groenwold, A.~A., \enquote{The Particle Swarm Optimization
  Algorithm in Size and Shape Optimization,} \emph{Structural and
  Multidisciplinary Optimization}, Vol.~23, 2002, pp. 259--267.
\newblock \doi{10.1007/s00158-002-0188-0}.

\bibitem[{Nejat et~al.(2014)Nejat, Mirzabeygi, and
  Shariat~Panahi}]{nejat2014airfoil}
Nejat, A., Mirzabeygi, P., and Shariat~Panahi, M., \enquote{Airfoil Shape
  Optimization Using Improved Multiobjective Territorial Particle Swarm
  Algorithm with the Objective of Improving Stall Characteristics,}
  \emph{Structural and Multidisciplinary Optimization}, Vol.~49, 2014, pp.
  953--967.
\newblock \doi{10.1007/s00158-013-1025-3}.

\bibitem[{Huang et~al.(2004)Huang, Huang, LeBeau, and
  Hauser}]{Huang2004numerical}
Huang, L., Huang, P., LeBeau, R.~P., and Hauser, T., \enquote{{Numerical Study
  of Blowing and Suction Control Mechanism on NACA0012 Airfoil},} \emph{Journal
  of Aircraft}, Vol.~41, No.~5, 2004, pp. 1005--1013.
\newblock \doi{10.2514/1.2255}.

\bibitem[{Yousefi and Saleh(2015)}]{Yousefi2015three}
Yousefi, K., and Saleh, R., \enquote{{Three-Dimensional Suction Flow Control
  and Suction Jet Length Optimization of NACA 0012 Wing},} \emph{Meccanica},
  Vol.~50, 2015, pp. 1481--1494.
\newblock \doi{10.1007/s11012-015-0100-9}.

\bibitem[{Munday and Taira(2014)}]{Munday2014separation}
Munday, P.~M., and Taira, K., \enquote{{Separation Control on NACA 0012 Airfoil
  Using Momentum and Wall-Normal Vorticity Injection},} \emph{32nd AIAA Applied
  Aerodynamics Conference}, 2014, p. 2685.
\newblock \doi{10.2514/6.2014-2685}.

\bibitem[{Greenblatt and Wygnanski(2000)}]{Greenblatt2000the}
Greenblatt, D., and Wygnanski, I.~J., \enquote{The Control of Flow Separation
  by Periodic Excitation,} \emph{Progress in Aerospace Sciences}, Vol.~36,
  No.~7, 2000, pp. 487--545.
\newblock \doi{10.1016/S0376-0421(00)00008-7}.

\bibitem[{{De Giorgi} et~al.(2015){De Giorgi}, {De Luca}, Ficarella, and
  Marra}]{DEGIORGI2015comparison}
{De Giorgi}, M., {De Luca}, C., Ficarella, A., and Marra, F.,
  \enquote{Comparison between Synthetic Jets and Continuous Jets for Active
  Flow Control: Application on a NACA 0015 and a Compressor Stator Cascade,}
  \emph{Aerospace Science and Technology}, Vol.~43, 2015, pp. 256--280.
\newblock \doi{10.1016/j.ast.2015.03.004}.

\bibitem[{Skarolek and {J. Karabelas}(2016)}]{SKAROLEK2016700}
Skarolek, V., and {J. Karabelas}, S., \enquote{Energy efficient Active Control
  of the Flow Past an Aircraft Wing: RANS and LES evaluation,} \emph{Applied
  Mathematical Modelling}, Vol.~40, No.~2, 2016, pp. 700--725.
\newblock \doi{10.1016/j.apm.2015.09.028}.

\bibitem[{Montazer et~al.(2016)Montazer, Mirzaei, Salami, Ward, Romli, and
  Kazi}]{Montazer2016Optimization}
Montazer, E., Mirzaei, M., Salami, E., Ward, T.~A., Romli, F.~I., and Kazi,
  S.~N., \enquote{Optimization of a Synthetic Jet Actuator for Flow Control
  Around an Airfoil,} \emph{IOP Conference Series: Materials Science and
  Engineering}, Vol. 152, No.~1, 2016, p. 012023.
\newblock \doi{10.1088/1757-899X/152/1/012023}.

\bibitem[{Smith and Swift(2003)}]{Smith2003comparison}
Smith, B.~L., and Swift, G.~W., \enquote{A Comparison between Synthetic Jets
  and Continuous Jets,} \emph{Experiments in Fluids}, Vol.~34, 2003, pp.
  467--472.
\newblock \doi{10.1007/s00348-002-0577-6}.

\bibitem[{Wu et~al.(1998)Wu, Lu, Denny, Fan, and Wu}]{Wu1998post}
Wu, J.-Z., Lu, X.-Y., Denny, A.~G., Fan, M., and Wu, J.-M., \enquote{Post-Stall
  Flow Control on an Airfoil by Local Unsteady Forcing,} \emph{Journal of Fluid
  Mechanics}, Vol. 371, 1998, p. 21–58.
\newblock \doi{10.1017/S0022112098002055}.

\bibitem[{Itsariyapinyo and Sharma(2022)}]{itsariyapinyo2022experimental}
Itsariyapinyo, P., and Sharma, R.~N., \enquote{Experimental Study of a NACA
  0015 Circulation Control Airfoil Using Synthetic Jet Actuation,} \emph{AIAA
  Journal}, Vol.~60, No.~3, 2022, pp. 1612--1629.
\newblock \doi{10.2514/1.J060508}.

\bibitem[{Tousi et~al.(2021)Tousi, Coma, Bergadà, Pons-Prats, Mellibovsky, and
  Bugeda}]{Tousi2021active}
Tousi, N., Coma, M., Bergadà, J., Pons-Prats, J., Mellibovsky, F., and Bugeda,
  G., \enquote{{Active Flow Control Optimisation on SD 7003 Airfoil at Pre and
  Post-Stall Angles of Attack Using Synthetic Jets},} \emph{Applied
  Mathematical Modelling}, Vol.~98, 2021, pp. 435--464.
\newblock \doi{10.1016/j.apm.2021.05.016}.

\bibitem[{Tousi et~al.(2022)Tousi, Bergadà, and Mellibovsky}]{TOUSI2022large}
Tousi, N., Bergadà, J., and Mellibovsky, F., \enquote{Large Eddy Simulation of
  Optimal Synthetic Jet Actuation on a SD 7003 Airfoil in Post-Stall
  Conditions,} \emph{Aerospace Science and Technology}, Vol. 127, 2022, p.
  107679.
\newblock \doi{10.1016/j.ast.2022.107679}.

\bibitem[{Zha and Paxton(2004)}]{Zha2004a}
Zha, G., and Paxton, C., \enquote{A Novel Airfoil Circulation Augment Flow
  Control Method Using Co-Flow Jet,} \emph{2nd AIAA Flow Control Conference},
  2004, p. 2208.
\newblock \doi{10.2514/6.2004-2208}.

\bibitem[{Zha et~al.(2006)Zha, Paxton, Conley, Wells, and
  Carroll}]{Zha2006effect}
Zha, G., Paxton, C.~D., Conley, C.~A., Wells, A., and Carroll, B.~F.,
  \enquote{Effect of Injection Slot Size on the Performance of Coflow Jet
  Airfoil,} \emph{Journal of Aircraft}, Vol.~43, No.~4, 2006, pp. 987--995.
\newblock \doi{10.2514/1.16999}.

\bibitem[{Vigneswaran and Kumar~GC(2021)}]{vigneswaran2021aerodynamic}
Vigneswaran, C., and Kumar~GC, V., \enquote{Aerodynamic Performance Analysis of
  Co-Flow Jet Airfoil,} \emph{International Journal of Aviation, Aeronautics,
  and Aerospace}, Vol.~8, No.~1, 2021, p.~10.
\newblock \doi{10.15394/ijaaa.2021.1555}.

\bibitem[{Sundaram et~al.(2022)Sundaram, Sengupta, Suman, Sengupta, Bhumkar,
  and Mathpal}]{Sundaram2022flow}
Sundaram, P., Sengupta, S., Suman, V.~K., Sengupta, T.~K., Bhumkar, Y.~G., and
  Mathpal, R.~K., \enquote{Flow Control Using Single Dielectric Barrier
  Discharge Plasma Actuator for Flow over Airfoil,} \emph{Physics of Fluids},
  Vol.~34, No.~9, 2022, p. 095134.
\newblock \doi{10.1063/5.0107638}.

\bibitem[{Yu and Zheng(2020)}]{yu2020numerical}
Yu, H., and Zheng, J., \enquote{Numerical investigation of Control of Dynamic
  Stall over a NACA0015 Airfoil Using Dielectric Barrier Discharge Plasma
  Actuators,} \emph{Physics of Fluids}, Vol.~32, No.~3, 2020.
\newblock \doi{10.1063/1.5142465}.

\bibitem[{Kim and Kim(2020)}]{KIM2020Effects}
Kim, S., and Kim, K., \enquote{Effects of Installation Location of Fluidic
  Oscillators on Aerodynamic Performance of an Airfoil,} \emph{Aerospace
  Science and Technology}, Vol.~99, 2020, p. 105735.
\newblock \doi{10.1016/j.ast.2020.105735}.

\bibitem[{Jones et~al.(2017)Jones, Milholen, Chan, and
  Goodliff}]{jones2017sweeping}
Jones, G.~S., Milholen, W.~E., Chan, D.~T., and Goodliff, S.~L., \enquote{A
  Sweeping Jet Application on a High Reynolds Number Semi-Span Supercritical
  Wing Configuration,} \emph{35th AIAA Applied Aerodynamics Conference}, 2017,
  p. 3044.
\newblock \doi{10.2514/6.2017-3044}.

\bibitem[{Koklu and Owens(2014)}]{koklu2014flow}
Koklu, M., and Owens, L.~R., \enquote{Flow Separation Control Over a Ramp Using
  Sweeping Jet Actuators,} \emph{7th AIAA Flow Control Conference}, 2014, p.
  2367.
\newblock \doi{10.2514/6.2014-2367}.

\bibitem[{Duriez et~al.(2017)Duriez, Brunton, and Noack}]{duriez2017machine}
Duriez, T., Brunton, S.~L., and Noack, B.~R., \emph{Machine Learning
  Control-taming Nonlinear Dynamics and Turbulence}, Vol. 116, Springer, 2017.
\newblock \doi{10.1007/978-3-319-40624-4}.

\bibitem[{Wang et~al.(2022)Wang, Mei, Aubry, Chen, Wu, and Wu}]{Wang2022deep}
Wang, Y., Mei, Y., Aubry, N., Chen, Z., Wu, P., and Wu, W., \enquote{Deep
  Reinforcement Learning Based Synthetic Jet Control on Disturbed Flow over
  Airfoil,} \emph{Physics of Fluids}, Vol.~34, No.~3, 2022, p. 033606.
\newblock \doi{10.1063/5.0080922}.

\bibitem[{Tadjfar and Asgari(2017)}]{Tadjfar2017active}
Tadjfar, M., and Asgari, E., \enquote{Active Flow Control of Dynamic Stall by
  Means of Continuous Jet Flow at Reynolds Number of $1\times 10^6$,}
  \emph{Journal of Fluids Engineering}, Vol. 140, No.~1, 2017, p. 011107.
\newblock \doi{10.1115/1.4037841}.

\bibitem[{Li et~al.(2022)Li, Chang, Kong, and Bao}]{Li2022recent}
Li, Y., Chang, J., Kong, C., and Bao, W., \enquote{Recent Progress of Machine
  Learning in Flow Modeling and Active Flow Control,} \emph{Chinese Journal of
  Aeronautics}, Vol.~35, No.~4, 2022, pp. 14--44.
\newblock \doi{10.1016/j.cja.2021.07.027}.

\bibitem[{Lee et~al.(1997)Lee, Kim, Babcock, and Goodman}]{lee1997application}
Lee, C., Kim, J., Babcock, D., and Goodman, R., \enquote{Application of Neural
  Networks to Turbulence Control for Drag Reduction,} \emph{Physics of Fluids},
  Vol.~9, No.~6, 1997, pp. 1740--1747.
\newblock \doi{10.1063/1.869290}.

\bibitem[{Garnier et~al.(2021)Garnier, Viquerat, Rabault, Larcher, Kuhnle, and
  Hachem}]{garnier2021review}
Garnier, P., Viquerat, J., Rabault, J., Larcher, A., Kuhnle, A., and Hachem,
  E., \enquote{A Review on Deep Reinforcement Learning for Fluid Mechanics,}
  \emph{Computers \& Fluids}, Vol. 225, 2021, p. 104973.
\newblock \doi{10.1016/j.compfluid.2021.104973}.

\bibitem[{Liu and MacArt(2024)}]{liu2024adjoint}
Liu, X., and MacArt, J.~F., \enquote{Adjoint-Based Machine Learning for Active
  Flow Control,} \emph{Physical Review Fluids}, Vol.~9, No.~1, 2024, p. 013901.
\newblock \doi{10.1103/PhysRevFluids.9.013901}.

\bibitem[{Sonoda et~al.(2023)Sonoda, Liu, Itoh, and
  Hasegawa}]{sonoda2023reinforcement}
Sonoda, T., Liu, Z., Itoh, T., and Hasegawa, Y., \enquote{Reinforcement
  Learning of Control Strategies for Reducing Skin Friction Drag in a Fully
  Developed Turbulent Channel Flow,} \emph{Journal of Fluid Mechanics}, Vol.
  960, 2023, p. A30.
\newblock \doi{10.1017/jfm.2023.147}.

\bibitem[{Lee et~al.(2023)Lee, Kim, and Lee}]{lee2023turbulence}
Lee, T., Kim, J., and Lee, C., \enquote{Turbulence Control for Drag Reduction
  through Deep Reinforcement Learning,} \emph{Physical Review Fluids}, Vol.~8,
  No.~2, 2023, p. 024604.
\newblock \doi{10.1103/PhysRevFluids.8.024604}.

\bibitem[{Portal-Porras et~al.(2023)Portal-Porras, Fernandez-Gamiz, Zulueta,
  Garcia-Fernandez, and {Etxebarria Berrizbeitia}}]{PortalPorras2023Active}
Portal-Porras, K., Fernandez-Gamiz, U., Zulueta, E., Garcia-Fernandez, R., and
  {Etxebarria Berrizbeitia}, S., \enquote{Active Flow Control on Airfoils by
  Reinforcement Learning,} \emph{Ocean Engineering}, Vol. 287, 2023, p. 115775.
\newblock \doi{10.1016/j.oceaneng.2023.115775}.

\bibitem[{Li(2018)}]{li2017deep}
Li, Y., \enquote{Deep Reinforcement Learning: An Overview,} \emph{arXiv
  preprint}, 2018.
\newblock \doi{10.48550/arXiv.1701.07274}.

\bibitem[{Duraisamy(2021)}]{Duraisamy2020perspectives}
Duraisamy, K., \enquote{Perspectives on Machine Learning-Augmented
  Reynolds-Averaged and Large Eddy Simulation Models of Turbulence,}
  \emph{Physical Review Fluids}, Vol.~6, No.~5, 2021, p. 050504.
\newblock \doi{10.1103/PhysRevFluids.6.050504}.

\bibitem[{Sirignano et~al.(2020)Sirignano, MacArt, and
  Freund}]{Sirignano2020DPM}
Sirignano, J., MacArt, J.~F., and Freund, J.~B., \enquote{{DPM: A Deep Learning
  PDE Augmentation Method with Application to Large-Eddy Simulation},}
  \emph{Journal of Computational Physics}, Vol. 423, 2020, p. 109811.
\newblock \doi{10.1016/j.jcp.2020.109811}.

\bibitem[{MacArt et~al.(2021)MacArt, Sirignano, and
  Freund}]{MacArt2021Embedded}
MacArt, J.~F., Sirignano, J., and Freund, J.~B., \enquote{Embedded Training of
  Neural-Network Subgrid-Scale Turbulence Models,} \emph{Physical Review
  Fluids}, Vol.~6, 2021, p. 050502.
\newblock \doi{10.1103/PhysRevFluids.6.050502}.

\bibitem[{Sirignano and MacArt(2023)}]{sirignano2023bluff}
Sirignano, J., and MacArt, J.~F., \enquote{Deep Learning Closure Models for
  Large-Eddy Simulation of Flows Around Bluff Bodies,} \emph{Journal of Fluid
  Mechanics}, Vol. 966, 2023, p. A26.
\newblock \doi{10.1017/jfm.2023.446}.

\bibitem[{Hickling et~al.(2024)Hickling, Sirignano, and
  MacArt}]{hickling2024large}
Hickling, T., Sirignano, J., and MacArt, J.~F., \enquote{Large Eddy Simulation
  of Airfoil Flows Using Adjoint-Trained Deep Learning Closure Models,}
  \emph{AIAA SciTech Forum}, 2024, p. 0296.
\newblock \doi{10.2514/6.2024-0296}.

\bibitem[{Sirignano et~al.(2023)Sirignano, MacArt, and
  Spiliopoulos}]{sirignano2023pde}
Sirignano, J., MacArt, J., and Spiliopoulos, K., \enquote{{PDE-Constrained
  Models with Neural Network Terms: Optimization and Global Convergence},}
  \emph{Journal of Computational Physics}, Vol. 481, 2023, p. 112016.
\newblock \doi{10.1016/j.jcp.2023.112016}.

\bibitem[{Nair et~al.(2023)Nair, Sirignano, Panesi, and MacArt}]{nair2023deep}
Nair, A.~S., Sirignano, J., Panesi, M., and MacArt, J.~F., \enquote{Deep
  Learning Closure of the Navier--Stokes Equations for Transition-Continuum
  Flows,} \emph{AIAA Journal}, Vol.~61, No.~12, 2023, pp. 5484--5497.
\newblock \doi{10.2514/1.J062935}.

\bibitem[{Kryger and MacArt(2024)}]{kryger2024optimization}
Kryger, M., and MacArt, J.~F., \enquote{{Optimization of Second-Order Transport
  Models for Transition-Continuum Flows},} \emph{AIAA Journal}, 2024.
\newblock \doi{10.48550/arXiv.2411.13515}.

\bibitem[{Lele(1992)}]{Lele1992}
Lele, S.~K., \enquote{Compact Finite Difference Schemes with Spectral-Like
  Resolution,} \emph{Journal of Computational Physics}, Vol. 103, No.~1, 1992,
  pp. 16--42.
\newblock \doi{10.1016/0021-9991(92)90324-R}.

\bibitem[{Ducros et~al.(1998)Ducros, Nicoud, and Poinsot}]{Ducros1998wale}
Ducros, F., Nicoud, F., and Poinsot, T., \enquote{Wall-Adapting Local
  Eddy-Viscosity Models for Simulations in Complex Geometries,} \emph{Numerical
  Methods for Fluid Dynamics VI}, Vol.~6, 1998, pp. 293--299.

\bibitem[{Jones et~al.(2008)Jones, Sandberg, and Sandham}]{jones2008direct}
Jones, L.~E., Sandberg, R.~D., and Sandham, N.~D., \enquote{Direct Numerical
  Simulations of Forced and Unforced Separation Bubbles on an Airfoil at
  Incidence,} \emph{Journal of Fluid Mechanics}, Vol. 602, 2008, p. 175–207.
\newblock \doi{10.1017/S0022112008000864}.

\bibitem[{Balakumar(2017)}]{Balakumar2017direct}
Balakumar, P., \enquote{Direct Numerical Simulation of Flows over an NACA-0012
  Airfoil at Low and Moderate Reynolds Numbers,} \emph{47th AIAA Fluid Dynamics
  Conference}, 2017, p. 3978.
\newblock \doi{10.2514/6.2017-3978}.

\bibitem[{Turner and Kim(2020{\natexlab{a}})}]{Turner2020effect}
Turner, J.~M., and Kim, J.~W., \enquote{Effect of Spanwise Domain Size on
  Direct Numerical Simulations of Airfoil Noise During Flow Separation and
  Stall,} \emph{Physics of Fluids}, Vol.~32, No.~6, 2020{\natexlab{a}}, p.
  065103.
\newblock \doi{10.1063/5.0009664}.

\bibitem[{Paszke et~al.(2019)Paszke, Gross, Massa, Lerer, Bradbury, Chanan,
  Killeen, Lin, Gimelshein, Antiga, Desmaison, Kopf, Yang, DeVito, Raison,
  Tejani, Chilamkurthy, Steiner, Fang, Bai, and Chintala}]{NEURIPS2019_9015}
Paszke, A., Gross, S., Massa, F., Lerer, A., Bradbury, J., Chanan, G., Killeen,
  T., Lin, Z., Gimelshein, N., Antiga, L., Desmaison, A., Kopf, A., Yang, E.,
  DeVito, Z., Raison, M., Tejani, A., Chilamkurthy, S., Steiner, B., Fang, L.,
  Bai, J., and Chintala, S., \enquote{{PyTorch: An Imperative Style,
  High-Performance Deep Learning Library},} \emph{Advances in Neural
  Information Processing Systems 32}, edited by H.~Wallach, H.~Larochelle,
  A.~Beygelzimer, F.~d\' Alch\'{e}-Buc, E.~Fox, and R.~Garnett, 2019, pp.
  8024--8035.

\bibitem[{Tieleman and Hinton(2012)}]{RMSProp}
Tieleman, T., and Hinton, G., \enquote{Lecture 6.5-RMSProp: Divide the Gradient
  by a Running Average of Its Recent Magnitude,} \emph{COURSERA: Neural
  Networks for Machine Learning}, Vol.~4, No.~2, 2012, pp. 26--31.

\bibitem[{Li et~al.(2011)Li, Sharma, and Arik}]{li2011energy}
Li, R., Sharma, R., and Arik, M., \enquote{Energy Conversion Efficiency of
  Synthetic Jets,} \emph{International Electronic Packaging Technical
  Conference and Exhibition}, 2011, pp. 115--122.
\newblock \doi{10.1115/IPACK2011-52034}.

\bibitem[{Girfoglio et~al.(2015)Girfoglio, Greco, Chiatto, and {de
  Luca}}]{Girfoglio2015Modelling}
Girfoglio, M., Greco, C.~S., Chiatto, M., and {de Luca}, L., \enquote{Modelling
  of Efficiency of Synthetic Jet Actuators,} \emph{Sensors and Actuators A:
  Physical}, Vol. 233, 2015, pp. 512--521.
\newblock \doi{10.1016/j.sna.2015.07.030}.

\bibitem[{Barrios et~al.(2023)Barrios, Ren, and Zha}]{Barrios2022Simulation}
Barrios, P.~A., Ren, Y., and Zha, G., \enquote{Simulation of 3D Co-Flow Jet
  Airfoil with Integrated Micro-Compressor Actuator at Different Cruise Mach
  Numbers,} \emph{AIAA SciTech Forum}, 2023, pp. 2023--2118.
\newblock \doi{10.2514/6.2023-2118}.

\bibitem[{Fatahian et~al.(2019)Fatahian, Lohrasbi~Nichkoohi, Salarian, and
  Khaleghinia}]{Fatahian2019comparative}
Fatahian, E., Lohrasbi~Nichkoohi, A., Salarian, H., and Khaleghinia, J.,
  \enquote{Comparative Study of Flow Separation Control Using Suction and
  Blowing over an Airfoil with/without Flap,} \emph{S{\=a}dhan{\=a}}, Vol.~44,
  No.~11, 2019, p. 220.
\newblock \doi{10.1007/s12046-019-1205-y}.

\bibitem[{Chen and Sun(2023)}]{Chen2023A}
Chen, W., and Sun, X., \enquote{A Comparative Study of the Influences of
  Leading-Edge Suction and Blowing on the Aerodynamic Performance of a
  Horizontal-Axis Wind Turbine,} \emph{Journal of Energy Engineering}, Vol.
  149, No.~1, 2023, p. 04022051.
\newblock \doi{10.1061/JLEED9.EYENG-4583}.

\bibitem[{Talnikar et~al.(2016)Talnikar, Wang, and
  Laskowski}]{talnikar2016unsteady}
Talnikar, C., Wang, Q., and Laskowski, G.~M., \enquote{Unsteady Adjoint of
  Pressure Loss for a Fundamental Transonic Turbine Vane,} \emph{Journal of
  Turbomachinery}, Vol. 139, No.~3, 2016, p. 031001.
\newblock \doi{10.1115/1.4034800}.

\bibitem[{Rodr{\'\i}guez et~al.(2013)Rodr{\'\i}guez, Lehmkuhl, Borrell, and
  Oliva}]{Rodriguez2013}
Rodr{\'\i}guez, I., Lehmkuhl, O., Borrell, R., and Oliva, A., \enquote{Direct
  numerical simulation of a NACA0012 in full stall,} \emph{International
  Journal of Heat and Fluid Flow}, Vol.~43, 2013, pp. 194--203.
\newblock \doi{10.1016/j.ijheatfluidflow.2013.05.002}.

\bibitem[{Turner and Kim(2020{\natexlab{b}})}]{Turner2020a}
Turner, J.~M., and Kim, J.~W., \enquote{Aerofoil dipole noise due to flow
  separation and stall at a low Reynolds number,} \emph{International Journal
  of Heat and Fluid Flow}, Vol.~86, 2020{\natexlab{b}}, p. 108715.
\newblock \doi{10.1016/j.ijheatfluidflow.2020.108715}.

\end{thebibliography}

\end{document}